\documentclass[a4paper,12pt,1p,sort&compress]{elsarticle}

\usepackage{setspace}
\usepackage[cdot,amssymb]{SIunits}
\usepackage[dvips]{epsfig}
\usepackage{amssymb}
\usepackage{upgreek}

\newcommand{\comment}[1]{}%\textcolor{red}{[#1]}}

\newcommand{\Idiff}[2]{\ensuremath{\mathrm{d}#1/\mathrm{d}#2}}

\newcommand{\eV}{\electronvolt}
\newcommand{\keV}{\kilo\eV}
\newcommand{\MeV}{\mega\eV}
\newcommand{\TeV}{\tera\eV}
\newcommand{\kV}{\kilo\volt}
\newcommand{\kVpm}{\kV\per\meter}
\newcommand{\MVpm}{\mega\volt\per\meter}

\newcommand{\muon}{\ensuremath{\upmu}}
\newcommand{\muplus}{\ensuremath{\muon^{+}}}
\newcommand{\muminus}{\ensuremath{\muon^{-}}}
\newcommand{\xray}{\textsc{x}-ray}
\newcommand{\xrays}{\xray s}
\newcommand{\alphaparticle}{\ensuremath{\upalpha}}
\newcommand{\alphaparticles}{\alphaparticle 's}

\newcommand{\americium}{\ensuremath{^{241}{\rm Am}}}
\newcommand{\iron}{\ensuremath{^{55}{\rm Fe}}}

\newcommand{\IdTds}{\Idiff{T}{s}}

\newcommand{\Istoppingpower}{\ensuremath{(1/\rho)\,\IdTds}}
\newcommand{\sub}[2]{\ensuremath{#1_{\mathrm{#2}}}}
\newcommand{\Teq}{\sub{T}{eq}}
\newcommand{\Tdep}{\sub{T}{dep}}

\newcommand{\MnKa}{Mn K$\upalpha$}

\newcommand{\SDD}{\textsc{sdd}}
\newcommand{\HV}{\textsc{hv}}
\newcommand{\PEEK}{\textsc{peek}}
\newcommand{\FCD}{\textsc{fcd}}
\newcommand{\MPP}{\textsc{mpp}}
\newcommand{\HLL}{\textsc{hll}}
\newcommand{\ADC}{\textsc{adc}}
\newcommand{\CFD}{\textsc{cfd}}
\newcommand{\FWHM}{\textsc{fwhm}}

\newcommand{\parfig}{figure}    %figure text in parentheses
\newcommand{\partab}{table}     %table text in parentheses
\newcommand{\parsec}{section}   %section text in parentheses
\newcommand{\textfig}{figure}   %figure text in open text
\newcommand{\textsec}{section}  %section text in open text

\begin{document}

\begin{frontmatter}

\title{Development of a Frictional Cooling Demonstration experiment}
\author[address:mpi]{Daniel Greenwald\footnote{Corresponding author, deg@mpp.mpg.de}}
\author[address:mpi,address:ihep]{Yu Bao}
\author[address:mpi]{Allen Caldwell}
\author[address:cern]{Daniel Koll\'{a}r}

\address[address:mpi]{Max Planck Institute for Physics, Munich, Germany}
\address[address:ihep]{Institute of High Energy Physics, Chinese Academy of Sciences, Beijing, China}
\address[address:cern]{CERN, Geneva, Switzerland}

\begin{abstract}
  A muon collider would open new frontiers of investigation in high
  energy particle physics, allowing precision measurements to be made
  at the \TeV\ energy frontier. One of the greatest challenges to
  constructing a muon collider is the preparation of a beam of muons
  on a timescale comparable to the lifetime of the muon. Frictional
  cooling is a potential solution to this problem. In this paper, we
  briefly describe frictional cooling and detail the Frictional
  Cooling Demonstration (\FCD) experiment at the Max Planck Institute
  for Physics, Munich. The \FCD\ experiment, which aims to verify the
  working principles behind frictional cooling, is at the end of the
  commissioning phase and will soon begin data taking.
\end{abstract}


\end{frontmatter}
\thispagestyle{plain}
\section{Introduction}
Since muons are over 200 times more massive than electrons, they could
be circularly accelerated to \TeV\ energies without the
synchrotron-radiation energy losses that electrons suffer. At the same
time, since, unlike hadrons, they are fundamental particles with no
constituent partons, their collision energies can be known to
comparitively low uncertainties. Muon colliders would therefore open
new frontiers of investigation in high energy particle physics,
allowing precision measurements to be made at TeV energies.

A wide range of physics can be studied at a \muplus\muminus\ collider
\cite{Ankenbrandt:1999as,Barger:jk}: A sub-\TeV\ collider can scan for
the $s$-channel resonance of the Higgs boson~\cite{Bar95} and the
thresholds for the production of pairs of light beyond-standard-model
particles \cite{Barger:rz,Lykken:qv,PhysRevLett.80.5489} with \MeV\
precision.  A \TeV\ collider can search for the heavy particles of a
new physics~\cite{Lykken:qv}. The high flux of muons at the front of
the collider would allow for high-precision muon physics studies, such
as searches for rare
decays~($\muon\rightarrow~e\gamma$,~$\muon\rightarrow e$). As well,
high energy muons can be used for high $Q^2$ deep inelastic scattering
with protons~\cite{schellman:166,cheung1998}.

One of the greatest challenges to constructing a muon collider is the
cooling of a beam of muons on a timescale comparable to the lifetime
of the muon.  Simulation of a muon collider front end utilizing
frictional cooling~\cite{Abramowicz:2004} indicated that such a scheme
is a viable option for producing high lumonsity \muplus\ and \muminus\
beams. The Muon Collider group of the Max Planck Institute for
Physics (\MPP), Munich, is commissioning the Frictional Cooling
Demonstration~(\FCD) experiment to verify the working principles
behind the scheme.

\section{Frictional Cooling}

\begin{figure}[!b]
  \includegraphics[width=\textwidth]{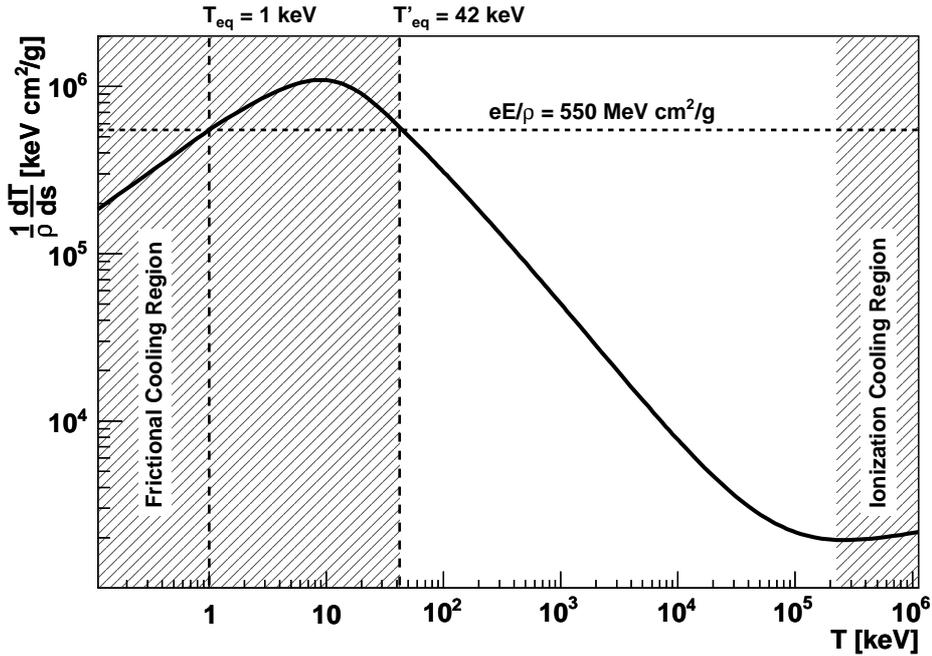}
  \caption{Stopping power (solid curve) of helium on \muplus. The
    horizontal dashed line shows the accelerating power of the
    restoring electric field for a particle of unit
    charge.\label{fig:dedx}}
\end{figure}

\begin{figure}[t]
  \begin{center}
  \includegraphics[width=\columnwidth]{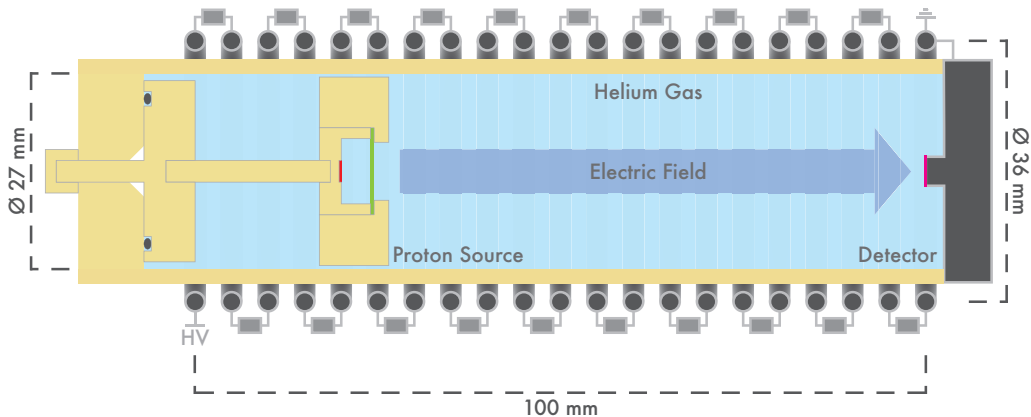}
  \caption{Scale diagram of the \FCD\ cooling
    cell.\label{fig:fcdCell}}
  \end{center}
\end{figure}

Frictional cooling involves the balancing of energy loss to a
moderator with energy gain from an electric field to bring a beam of
charged particles to an equilibrium energy and reduce dispersion. It
requires the beam be in an energy region where the stopping power of
the moderator medium---the energy loss per unit path length normalized
by the medium density, \Istoppingpower---increases with increasing
kinetic energy $T$.  There are two energy regions where this
requirement is met~(\parfig~\ref{fig:dedx}). Ionization cooling
schemes utilize particle beams in the high energy region~\cite{Neu83}.
Frictional cooling utilizes particle beams in the low energy region.%

Applying an electric field to restore energy loss creates two
equilibrium energies: a stable one at an energy \Teq\ below the
ionization peak of the stopping power curve, where the energy~loss
per unit~path~length $\IdTds\propto\sqrt{T}$, and an unstable one at
an energy $\Teq'$ above the peak, where \IdTds\ decreases with
increasing kinetic energy. Particles with kinetic energies below \Teq\
accelerate; those with kinetic energies between \Teq\ and $\Teq'$
decelerate. The coolable energy region is defined by $T<\Teq'$.
Additionally, restoring lost energy only in the longitudinal direction
provides transverse cooling.

For a chosen equilibrium energy, the electric field strength required
to balance the energy loss scales directly with the density of the
moderator. The density of the moderator must therefore be low, to keep
the electric field strength within a feasible range. Helium and
hydrogen gasses are good moderators because they have low densities
and suppress the capture of electrons by the cooled
particles~\cite{PhysRevLett.33.568} and the capture of the particles
by the medium atoms~\cite{Coh02} \comment{plus proton ref}.

\section{Frictional Cooling Demonstration Experiment\label{sec:construction}}

The \FCD\ experiment at the \MPP\ is designed to verify the working
principles behind frictional cooling and the modeling of frictional
cooling used in the simulation. The dependence of \Teq\ on moderator
density and electric field strength can be measured and compared to
monte carlo simulations.

\begin{figure}[t]
  \begin{center}
  \includegraphics[width=\columnwidth]{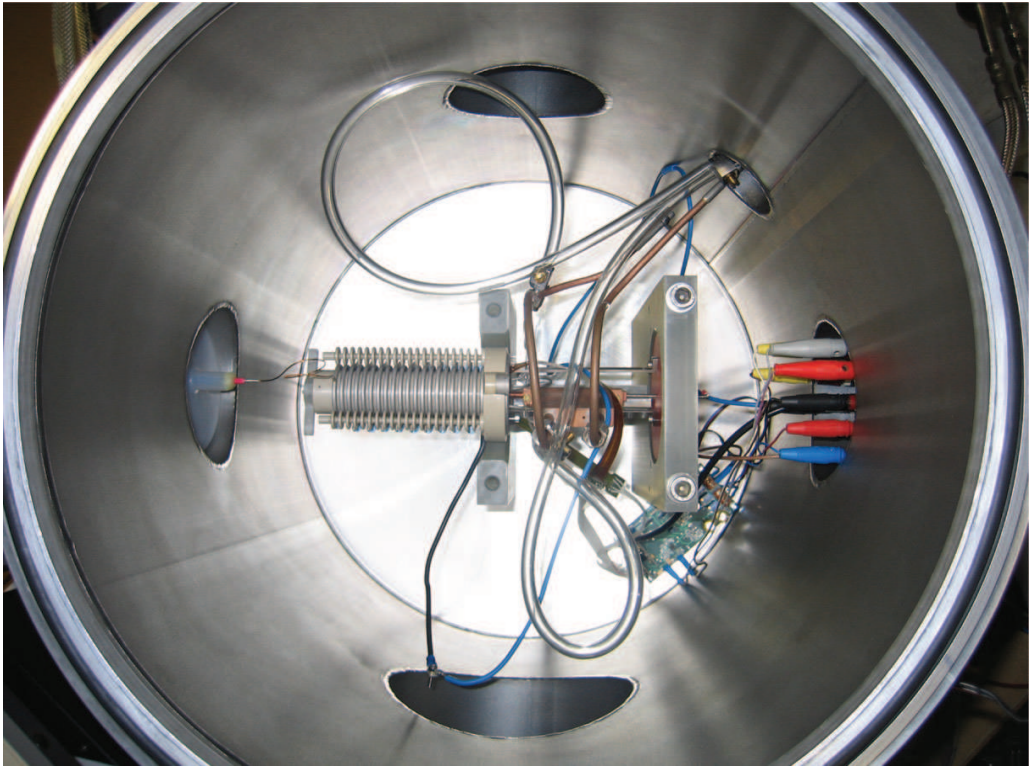}
  \caption{Photo of the experimental setup: visible are the gas cell
    (center) with the source on the left-hand side and detector on the
    right-hand side; the connection to \HV\ (left); electronics feedthroughs
    (right) leading to the electronics (bottom right); and the gas
    feedthroughs (top right).
    \label{fig:fcdphoto}}
  \end{center}
\end{figure}

The experiment consists of a gas cell mounted inside an
accelerating grid that provides the restoring electric
field~(\parfig~\ref{fig:fcdCell}). A proton source and an open silicon
drift detector~(\SDD) are mounted inside the gas cell. The whole
construction is then placed inside a vacuum tank
(\parfig~\ref{fig:fcdphoto}).

The accelerating grid is constructed from twenty-one metal
\comment{stainless-steel or tungsten} rings, \unit{3}{\milli\meter}
thick, spaced~\unit{5}{\milli\meter} apart, connected in series with
\unit{64}{\mega\ohm} resistors between rings. The rings enclose a
cylindrical space~\unit{33}{\milli\meter} in diameter
and~\unit{100}{\milli\meter} long from the center of the first ring to
the center of the last ring.  The first ring is connected to a power
supply capable of providing voltages up to~\unit{100}{\kilo\volt}; the
last ring is grounded.  The central axis of the grid defines the $z$
direction, with~$z\,=\,\unit{0}{\milli\meter}$ at the center of the
high-voltage (\HV) ring and~$z\,=\,\unit{100}{\milli\meter}$ at the
center of the ground ring. The grid creates a nearly uniform electric
field along $z$ with strengths up to \unit{1}{\mega\volt\per\meter}
(see
\parsec~\ref{sec:sim:efield}).

The gas cell is a cylinder made of \PEEK, with an outer radius of
\unit{31}{\milli\meter} and inner radius of \unit{27}{\milli\meter}.
It is centered inside the accelerating grid. The end of the cell
nearest the \HV\ ring is sealed but for a small hole on the $z$ axis
through which the proton source is mounted. The end of the cell
nearest the ground ring is sealed by a grounded metal flange that
holds the \SDD\ and provides gas input and output feedthroughs.

\begin{figure}[t]
  \begin{center}
    \includegraphics[width=0.45\columnwidth]{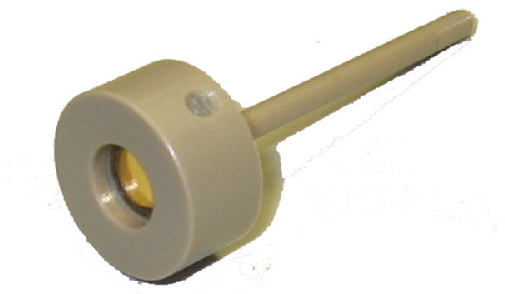}
    \includegraphics[width=0.45\columnwidth]{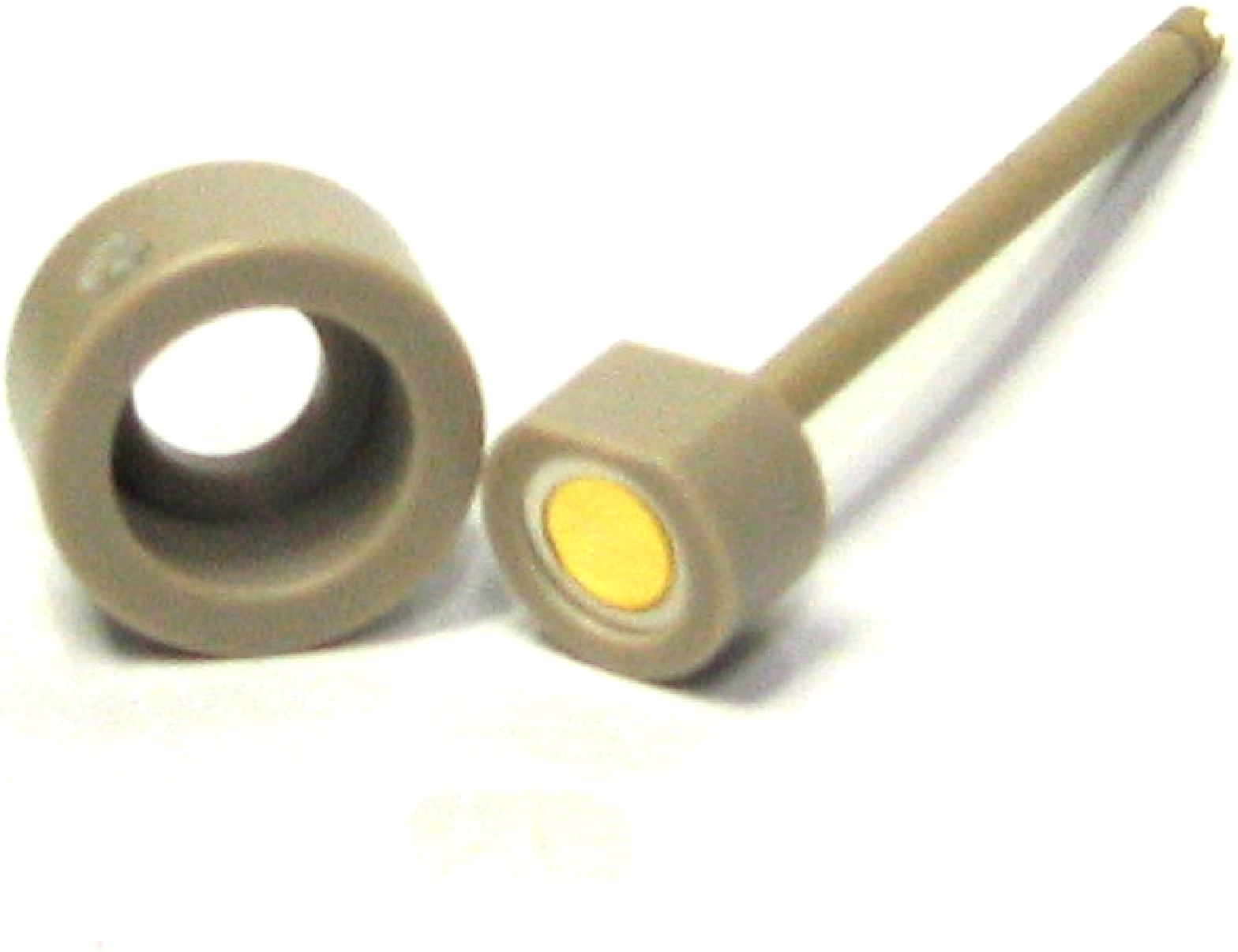}
    \caption{Photo of the proton source assembled (left) and
      disassembled (right). The gold-colored disc is the americium.
      The cylindrical cap holds the mylar
      foil.\label{fig:protonsourcephoto}}
  \end{center}
\end{figure}

The proton source (\parfig~\ref{fig:protonsourcephoto}) consists of an
open alpha source covered with a Mylar foil. The alpha particles free
protons from the Mylar (see \parsec~\ref{sec:sim:source}). The source
is embedded in the top of a lollipop made of \PEEK. A cap that fastens
to the lollipop head holds the Mylar foil in place. Foils of various
thicknesses can be swapped into the construction easily and quickly.

The lollipop stick screws into the head at one end and into a
cylindrical platform on the opposite end. This platform is
\unit{25}{\milli\meter} in diameter. The platform with the attached
lollipop fastens to the \HV\ end of the gas cell forming a gas-tight
seal.

\begin{figure}[t]
  \begin{minipage}[t]{0.45\textwidth}
    \centering
    \includegraphics[height=5.5cm]{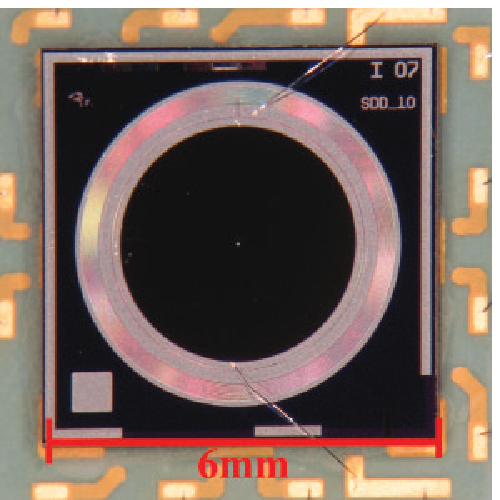}
    \caption{Photo of the \SDD. The inner black circle is the back
      surface of the active region.\label{fig:sddphoto}}
  \end{minipage}
  \begin{minipage}[t]{0.05\textwidth}
    \ 
  \end{minipage}
  \begin{minipage}[t]{0.5\textwidth}
    \centering
    \includegraphics[height=5.5cm]{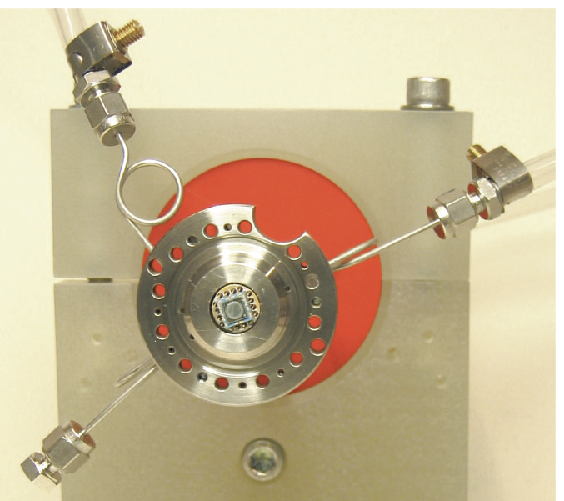}
    \caption{Photo of the \SDD\ mounted into the gas feedthrough flange.\label{fig:sddflangephoto}}
  \end{minipage}
\end{figure}

The \SDD\ (\parfig~\ref{fig:sddphoto}) mounts through the gas
feedthrough flange (\parfig~\ref{fig:sddflangephoto}) to a \PEEK\
holder; when fastened tightly, this mounting provides a gas-tight
seal. The \PEEK\ holder also acts as an electronics feedthrough for
the \SDD. This construction allows for a quick exchange of the \SDD.

\subsection{Electric Field \label{sec:sim:efield}}

\begin{figure}[t]
  \includegraphics[width=\columnwidth]{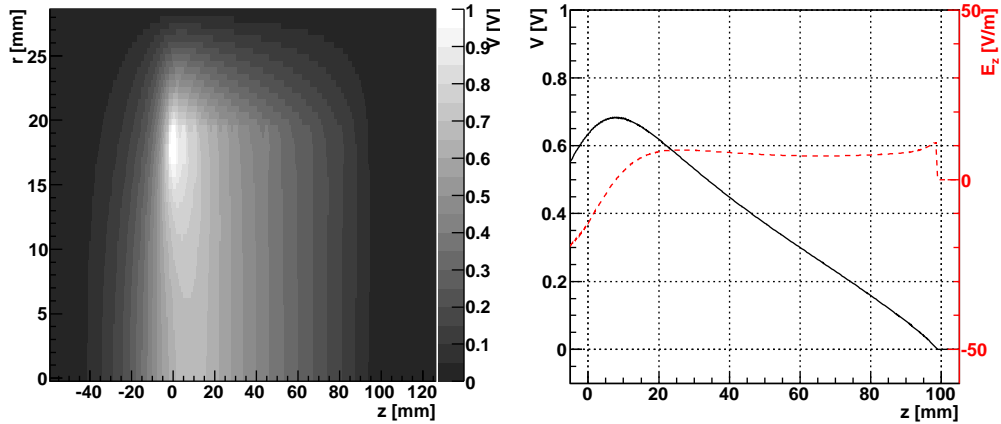}
  \caption{Map of the potential, for a voltage of \unit{1}{\volt}
    applied to the first ring, in the $z$--$r$ plane (left) and the
    potential (black; left axis) and longitudinal electric field
    strength (dashed; right axis) on the $z$ axis (right) created by
    the accelerating grid (with plastic source
    holder).\label{fig:efield}}
\end{figure}

A map of the electric field created by the accelerating grid is needed
for the full simulation of the \FCD\ experiment
(\parsec~\ref{sec:CoolSim}) as well as for characterizing the
detector's response to protons (\parsec~\ref{sec:protons}). We use a
successive overrelaxation algorithm to calculate the electric field
(\parfig~\ref{fig:efield}). This calculation revealed that the
potential is at its maximum not at $z=\unit{0}{\milli\meter}$, but
instead at $z=\unit{9}{\milli\meter}$. Therefore the surface of the
proton source must be placed at $z>\unit{9}{\milli\meter}$. The
electric field is strongest and also nearest to uniform at
$z>\unit{20}{\milli\meter}$. In the simulation of the \FCD\ cell and
in the experimental setup for the measurement of proton spectra, the
source surface is therefore placed at $z=\unit{20}{\milli\meter}$.

\subsection{Proton Source \label{sec:sim:source}}

\begin{figure}[t]
  \includegraphics[width=\columnwidth]{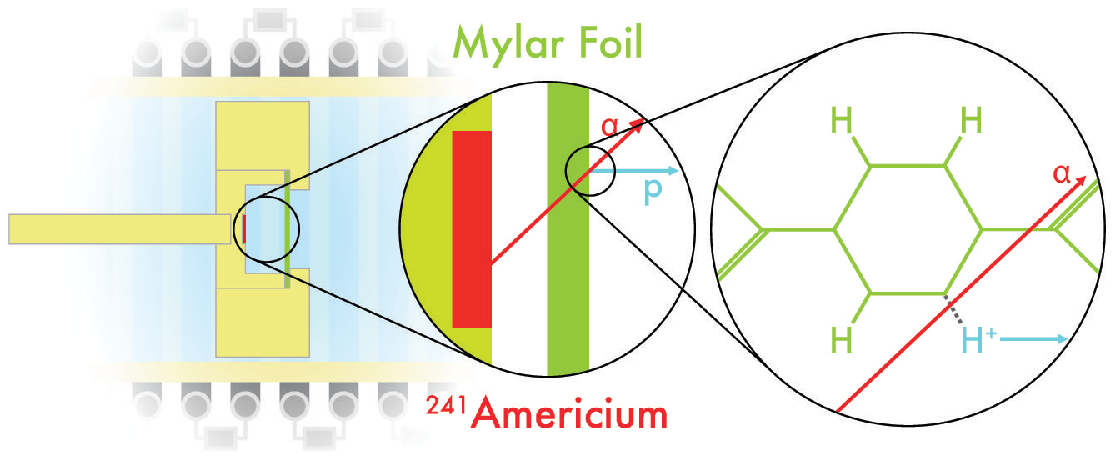}
  \caption{Schematic of the proton production mechanism \label{fig:prosrc}}
\end{figure}

\begin{table}[b]
  \begin{center}
    \begin{tabular}{l *{3}{r}}
      \hline
      \hline

      \rule{0pt}{1em}Energy (\MeV) & 5.388 & 5.422 & 5.485 \\
      Branching Ratio (\%)       & 1.0   & 13.0  & 84.5   \\

      \hline
      \hline
    \end{tabular}
  \end{center}

  \caption{Energies and branching ratios (BR) for alpha particles
    emitted by \americium\ with branching ratio greater than
    0.4\%\label{tab:AmAlphas}}	
\end{table}

The proton source contains a \unit{74}{\kilo\becquerel} \americium\
alpha source covered by a thin Mylar foil. The americium emits alpha
particles with energies approximately \unit{5}{\MeV}
(\partab~\ref{tab:AmAlphas}). As they pass through the Mylar, they
break carbon--hydrogen bonds, freeing hydrogen nuclei from the Mylar
molecule~(\parfig~\ref{fig:prosrc}). When these bonds are broken near
the outer surface of the foil, the electric field can accelerate the
resultant protons out of the foil before they are recaptured.

\begin{figure}[t]
  \includegraphics[width=\columnwidth]{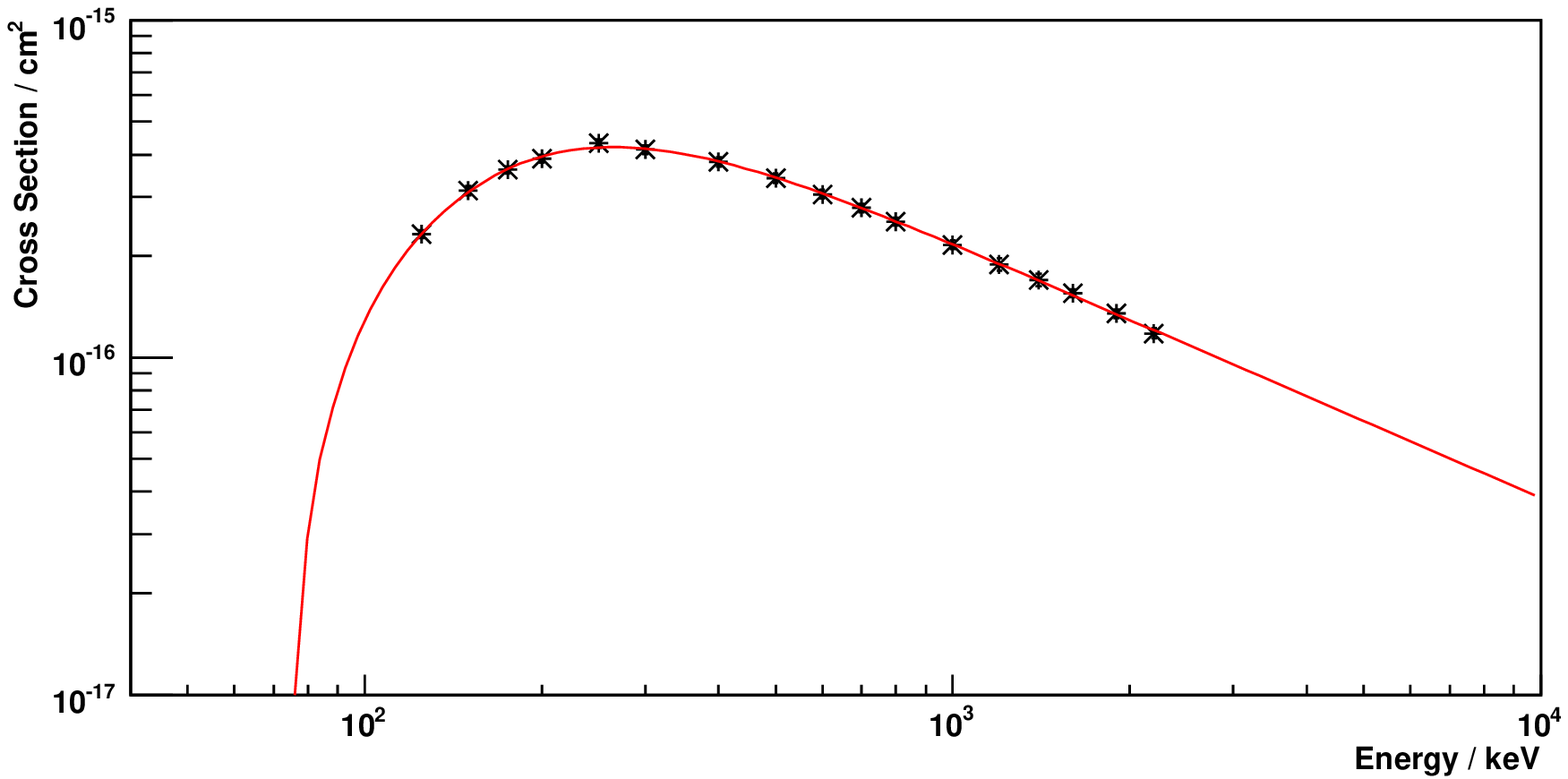}
  \caption{Experimentally measured cross section for the ionization of
    molecular hydrogen by He$^{2+}$ \cite{Tawara1985} (points) and a
    fit adapted from \cite{Gre71} (line) \label{fig:HeHCS}}
\end{figure}

The number of bonds an \alphaparticle\ breaks per unit distance is
{\sub{n}{p}(E) = \sub{\sigma}{ion}(E)\cdot\sub{\rho}{H}}, where
\sub{\sigma}{ion} is the cross section for ionization of molecular
hydrogen by He$^{2+}$ and \sub{\rho}{H} is the concentration of
hydrogen in Mylar, \unit{34.35}{\nano\meter\rpcubed}. The measured
cross section for ionization of molecular hydrogren,
\begin{displaymath}
  \textrm{He}^{2+} + \textrm{H}_{2} \to \textrm{He}^{n} + \textrm{p},\nonumber
\end{displaymath}
where He$^{n}$ is any charge state of helium, was reported in
\cite{Tawara1985}. To describe the cross section
(\parfig~\ref{fig:HeHCS}), we fit to the data a semi-empirical formula
for the cross section of hydrogen on helium found in \cite{Gre71},
\begin{displaymath}
  \sigma = \sigma_0 \cdot a_1 \, \left(\frac{E'}{E_R}\right)^{\displaystyle a_2}{\Bigg /}
  \left(1+\left(\frac{E'}{a_3}\right)^{\displaystyle a_2+a_4}
    + \left(\frac{E'}{a_5}\right)^{\displaystyle a_2+a_6}\right), \nonumber
\end{displaymath}
where $E'=E-E_t$ is the energy of the \alphaparticle\ minus the
threshold energy of the process, $E_t$; $E_R$ is the Rydberg energy
multiplied by $\sub{m}{He}/\sub{m}{e}$. The $a_i$ are the fit
parameters, and $\sigma_0$ is a scaling factor equal to
\unit{\power{10}{-16}}{\centi\meter\squared}.

The americium is in the form of a disc \unit{2.5}{\milli\meter} in
diameter, which is embedded in a plastic lollipop-shaped holder. It is
completely open on its exposed side. A cap fits over the lollipop to
hold the Mylar foil in place at a distance of
\unit{2.45}{\milli\meter} from the source. The cap has a circular
opening \unit{3.5}{\milli\meter} in diameter centered over the source.

\begin{figure}[t]
  \begin{center}
    \includegraphics[width=.45\columnwidth]{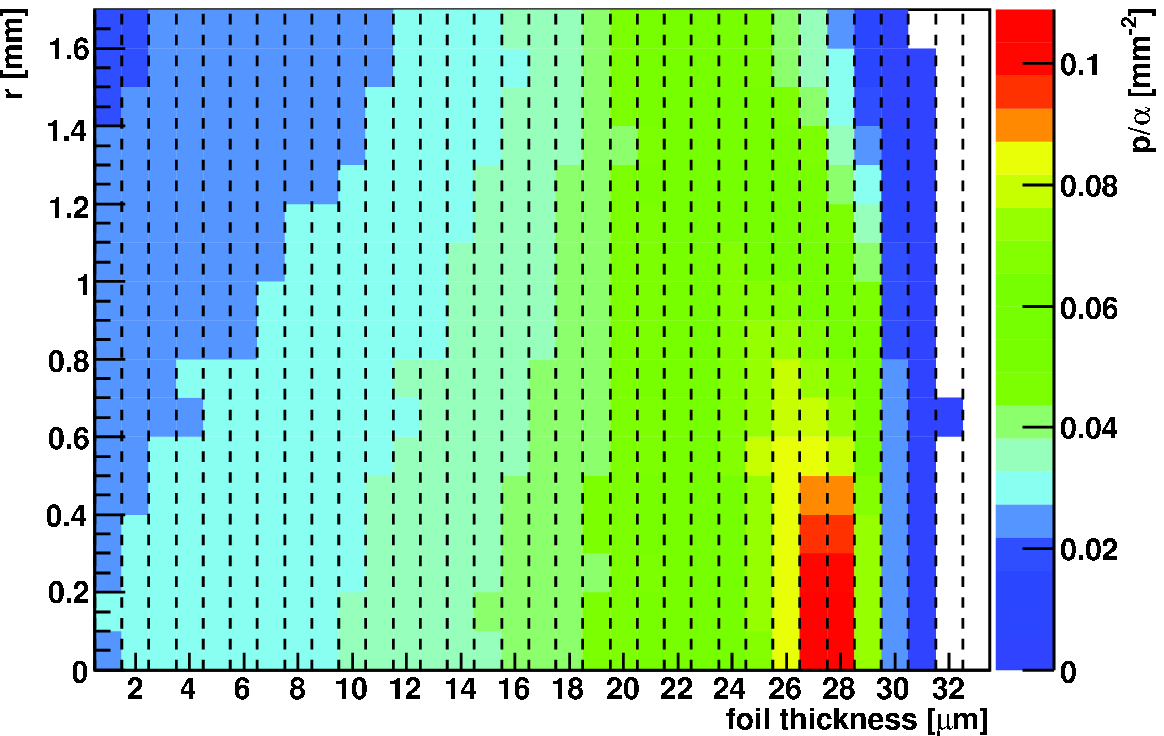}
    \includegraphics[width=.45\columnwidth]{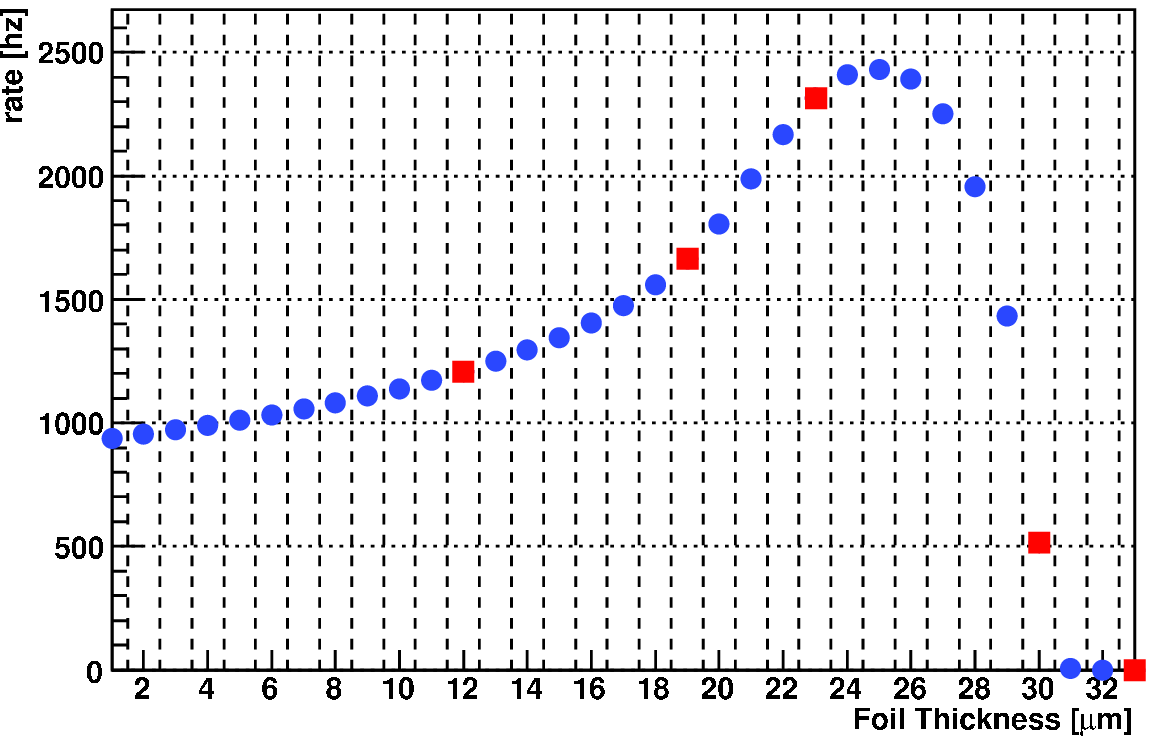}
    \caption{Radial distribution of p/\alphaparticle\ (left) and total
      proton production rate for a \unit{74}{\kilo\becquerel} alpha
      source (right) as functions of Mylar foil thickness. Both rates
      count only protons produced in the last \unit{1}{\nano\meter} of
      foil.\label{fig:prodist}}
    \end{center}
\end{figure}

We simulated this proton source in Geant4~\cite{Geant4}. The americium
emits alpha particles isotropically. We tracked those alpha particles
that pass through the foil and record their trajectories and energy
losses in the Mylar foil. We calculate the number of C--H bonds broken
by an \alphaparticle\ at a point along its trajectory using its
Geant4-calculated energy. This is used to calculate the radial
distribution (in a plane parallel to the surface of the foil) of the
number of C--H bonds broken within the last {1-\nano\meter} as a
function of thickness of the foil traversed. We used
\unit{1}{\nano\meter} as an estimate of the depth from which a proton
can escape out of the foil because this is roughtly the size of the
Mylar monomer.  Figure~\ref{fig:prodist} (left) displays the
results for thickness of the foil from \unit{1}{\micro\meter} to
\unit{33}{\micro\meter} in \unit{1}{\micro\meter} steps; this is the
precision on the thickness at which such foils are manufactured.

We also calculated the total proton production (again in the last
\unit{1}{\nano\meter}) as a function of thickness of the Mylar foil
(\parfig~\ref{fig:prodist}, right).  The shape of the
rate--thickness relationship is similar to that of the Bragg peak for
\alphaparticles\ in Mylar; however, it is broadened by the
distribution of the incidence angles of the \alphaparticles. The rate
is maximum in the region of \unit{23}{\micro\meter} and is zero above
\unit{30}{\micro\meter}, since the \alphaparticles\ are stopped in the
foil before reaching the surface. Around the rate-maximizing
thickness, the proton production is spatially uniform, and at larger
thicknesses, the protons are produced mainly at the center of the foil
surface (\parfig~\ref{fig:prodist}). Though at the larger thicknesses
the total rate is lower, when detector acceptance is taken into
account (see \parsec~\ref{sec:detAcc}), the centralized proton
production may be beneficial. The square-shaped markers in the right
plot of figure~\ref{fig:prodist} indicate the five thicknesses
available for the experiment: 12, 19, 24, 30, and
\unit{33}{\micro\meter}. By sandwiching foils together, we are also
able to test a thickness of \unit{31}{\micro\meter}.

\subsection{Silicon Drift Detector}

\begin{figure}[t]
  \begin{center}
    \includegraphics[width=0.5\columnwidth]{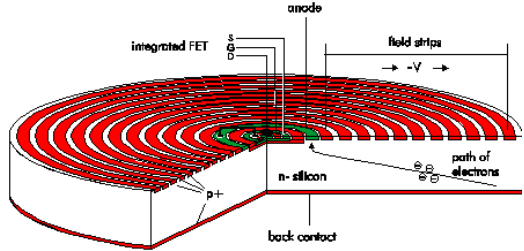}
    \caption{Schematic of the \SDD\ from \cite{Sim07}, showing the
      electrode structure used to deplete the detector bulk and setup
      the charge-collection field, and the
      \textsc{fet}.\label{fig:sddschematic}}
  \end{center}
\end{figure}

The \SDD, designed and constructed by the \MPP's Semiconductor
Laboratory, measures particle energies in the range \unit{100s}{eV} to
approximately \unit{150}{\keV} \comment{find \HLL ref, and fix
  numbers}. The detector is constructed on an n-type silicon wafer
\unit{450}{\micro\meter} thick. The exposed (back) surface of the
detector is covered uniformally with a 30-\nano\meter-thick aluminum
electrode.  The opposite surface is implanted with concentric rings of
p-type silicon (\parfig~\ref{fig:sddschematic}). A negative potential
(on the order of \unit{-100}{\volt}) on the aluminum depletes the
silicon.  The p-type rings are placed at voltages that produce a
well-shaped potential inside the silicon. Ionizing particles produce a
number of electron-hole pairs in the silicon in proportion to the
amount of energy they deposit along their trajectory; the electrons
then drift to the center of the p-doped surface of the detector, where
a field-effect transistor produces a signal.

\begin{figure}[t]
  \includegraphics[width=.5\columnwidth]{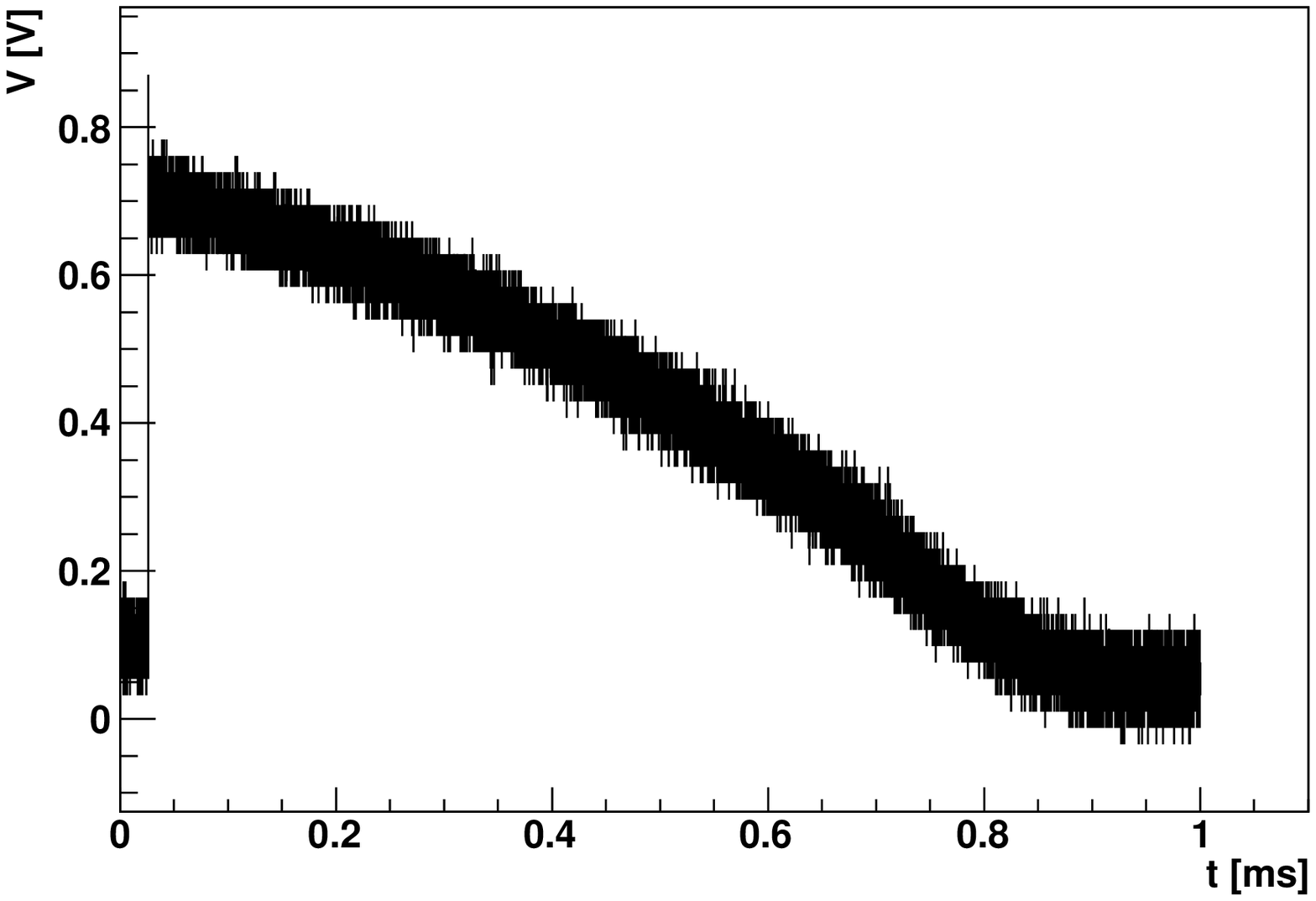}
  \includegraphics[width=.5\columnwidth]{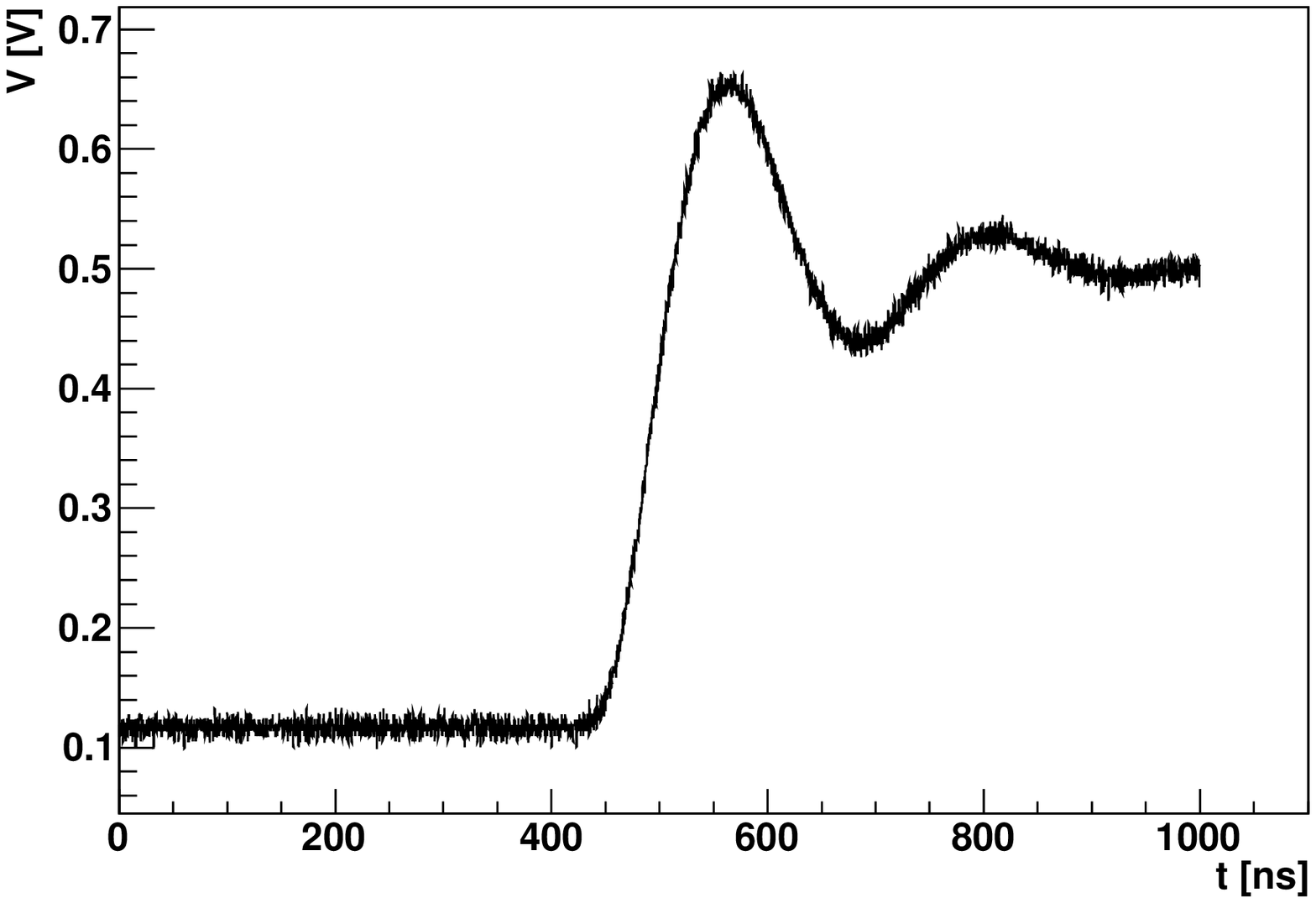}
  \caption{Signal from \SDD\ after initial amplification. The plot
    on the right shows the rising edge of the signal, shown in full on the left.
    \label{fig:preampSig}}
\end{figure}

The voltages for depletion of the silicon and setting up of the
potential well are regulated by electronics manufactured by
PNSensor~\cite{pnsensor}.  The detector outputs a
saw-tooth-shaped voltage pulse that after initial amplification by
these electronics has a rise time of 30~to~40~\nano\second\
(\parfig~\ref{fig:preampSig}) and an
amplitude~$\sub{V}{det}~\!\approx~\!(\unit{1.2}{\milli\volt\per\keV})\;\Tdep$,
where \Tdep\ is the energy the particle deposits in the active layers
of the detector.

\subsection{Electronic Readout}

\begin{figure}[t]
  \begin{minipage}[t]{0.5\textwidth}
    \centering
    \includegraphics[width=\columnwidth]{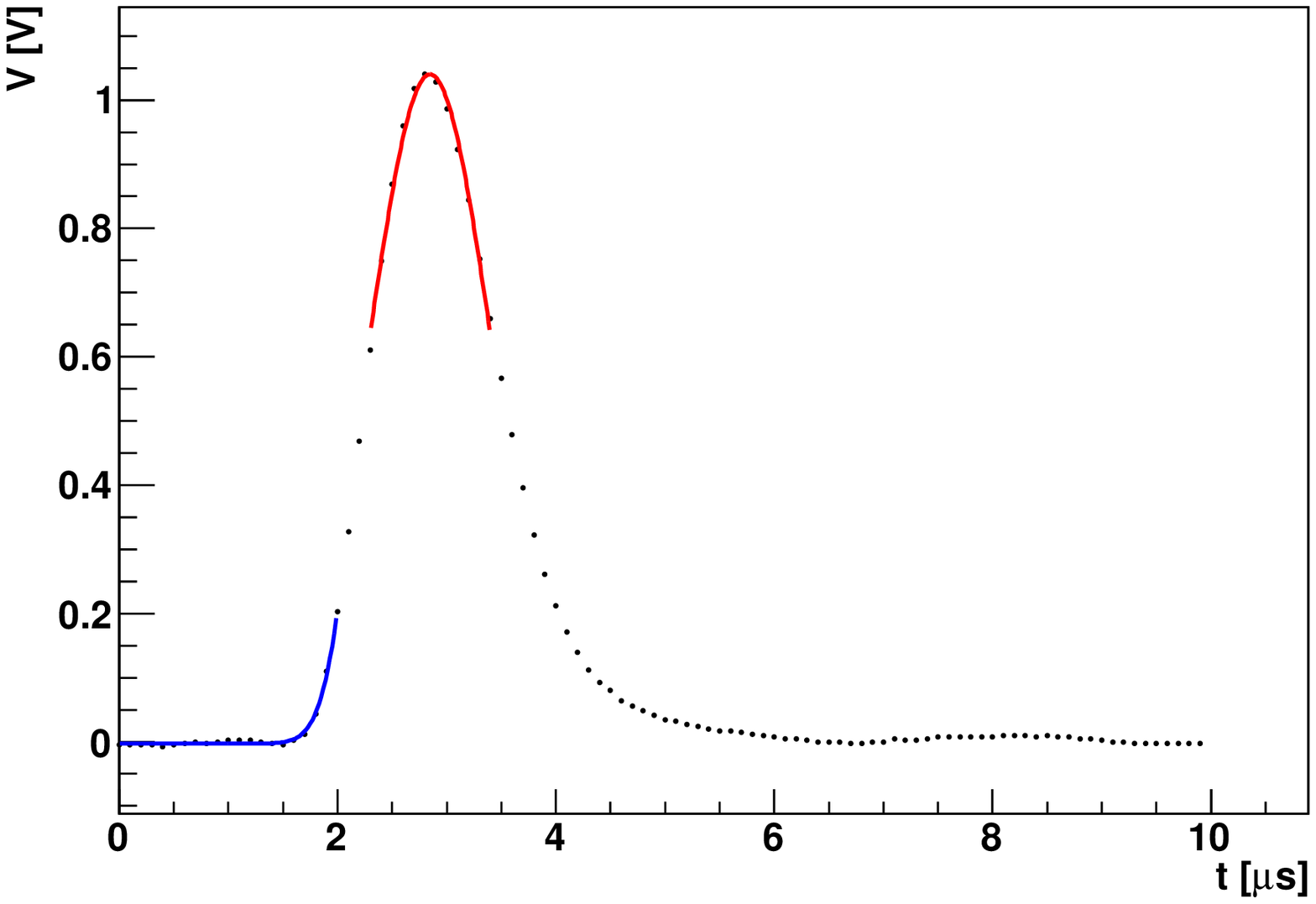}
  \end{minipage}
  \begin{minipage}[t]{0.5\textwidth}
    \centering
    \includegraphics[width=\textwidth]{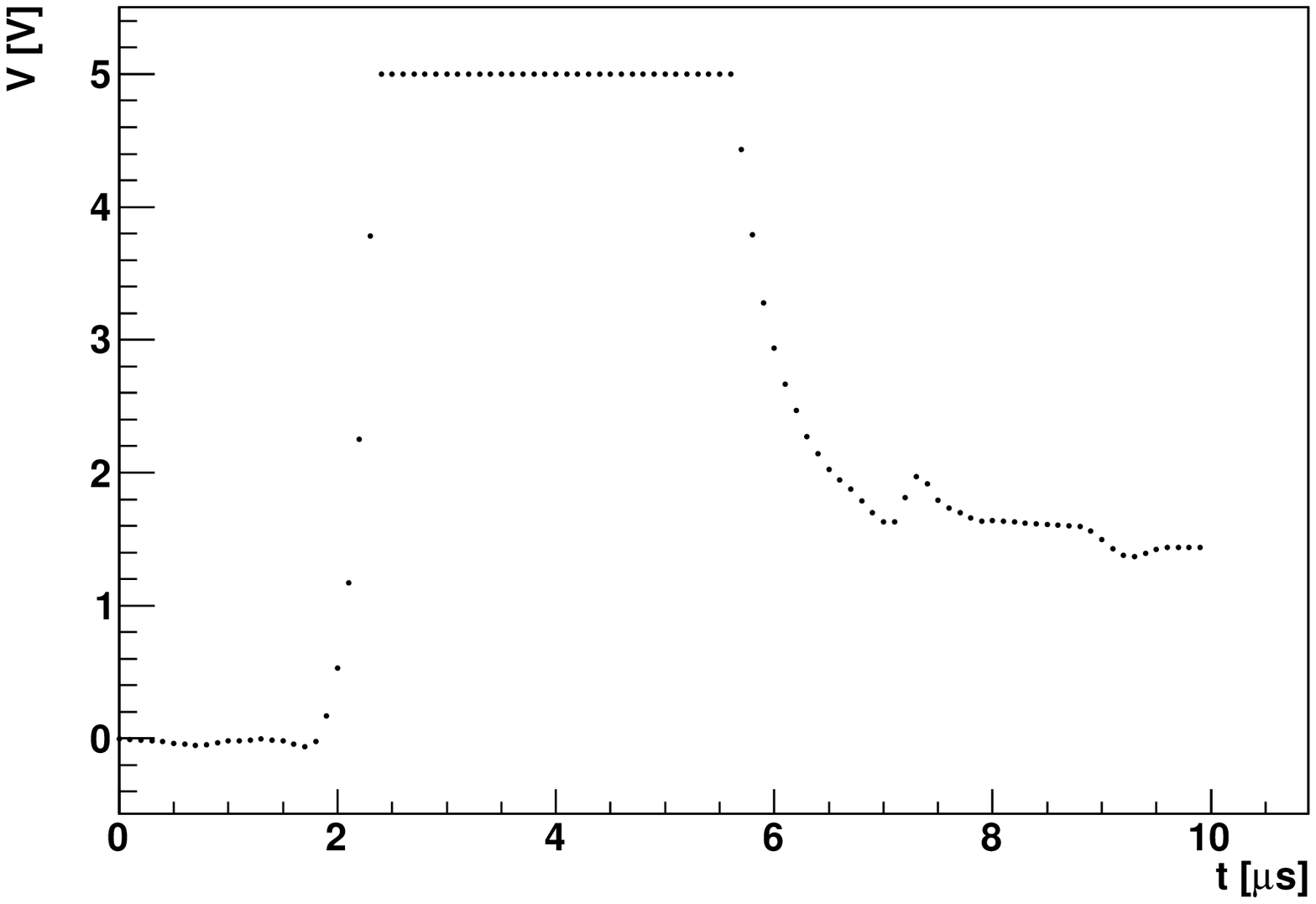}
  \end{minipage}

  \begin{minipage}[t]{0.47\textwidth}
    \caption{Shaped and digitized signal from the \SDD\ with pedestal and peak
      fits.\label{fig:shapedSig}\label{fig:fittedsig}}
  \end{minipage}
  \begin{minipage}[t]{0.03\textwidth}
    \ 
  \end{minipage}
  \begin{minipage}[t]{0.5\textwidth}
    \caption{Saturated signal from the \SDD.\label{fig:satSig}}
  \end{minipage}
\end{figure}

A shaping amplifier with a shaping time of~\unit{0.25}{\micro\second}
converts this quick signal to a fin-shaped pulse~3~to~4~\micro\second\
wide~(\parfig~\ref{fig:shapedSig}). The shaper preserves the linearity of
the signal amplitude's dependence on~\Tdep.  The signal is split in
two: one part is used for triggering; the other is digitized and saved
for offline analysis.

A 12-bit \ADC\ from National Instruments~\cite{natInst}, interfaced
with a computer via LabView~\cite{LabView}, digitizes the signal with
a sampling rate of \unit{10}{\mega\hertz}. Before entering the \ADC,
the signal is delayed \unit{4.75}{\micro\second} with respect to the
trigger signal, so that the baseline voltage before the signal is
recorded as well.  The \ADC\ records the event for
\unit{10}{\micro\second} (100 samples).

A window discriminator, consisting of two trailing-edge constant
fraction discriminators (\CFD), one setting a low threshold, the other
a high threshold, produces the trigger signal for the \ADC\ to begin
sampling. The low-threshold \CFD\ filters out low-amplitude voltage
fluctuations---that is, noise. Since the \ADC\ has a maximum
triggering rate of \unit{30}{\hertz}, when needed, the high-threshold
\CFD\ was used to filter out unimportant signals that had large
amplitudes, namely those from \alphaparticle\ particles.

A particle that deposits too much energy in the detector produces a
charge-saturated signal~(\parfig~\ref{fig:satSig}). The signal is
distorted and the energy of the particle cannot be reconstructed.
These signals themselves are filtered out by the window discriminator;
however, the saturation often produces secondary signal peaks in the
trailing edge of the original signal. These peaks are large enough to
pass the low-threshold \CFD\ but small enough for the high-threshold
\CFD\ to not veto them. To prevent these signals from swamping the
\ADC, a gate generator can produce a veto signal from the
high-threshold \CFD's trigger pulse. The length of the gate can be set
within a large range of times from less than~\unit{100}{\nano\second}
to greater than~\unit{11}{\second}.

\subsection{Offline Analysis}

The amplitude of a signal above the baseline linearly corresponds to
the energy deposited in the detector by the incoming particle. The
shape of the signal around its peak is approximately gaussian.
However, since outside this region the signal is not perfectly
gaussian, we cannot fit the whole signal with a gaussian shape plus a
pedestal to get both the amplitude and the baseline. Instead, we fit
the peak and the samples before the signal
separately~(\parfig~\ref{fig:fittedsig}). Fitting the first
approximately \unit{2}{\micro\second} with a constant pedestal plus a
gaussian function that describes the start of the signal gives the
baseline value. Fitting a symmetric \unit{2}{\micro\second} window
around the peak with a gaussian function gives the signal amplitude.

\begin{figure}[t]
  \includegraphics[width=\columnwidth]{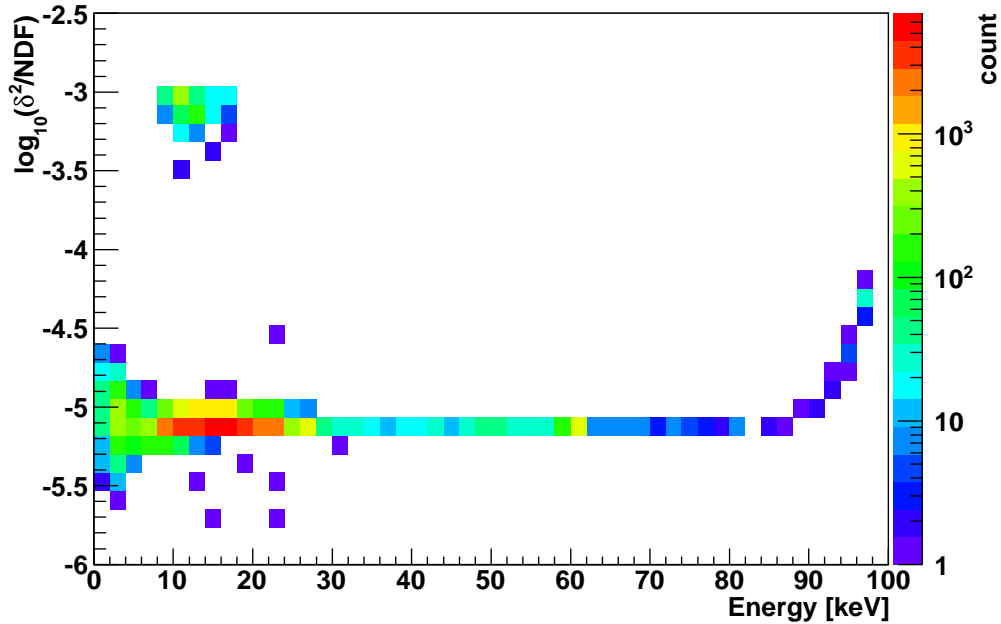}
  \caption{$\chi^{2}/$NDF distributions as a function of event energy
    for a background run.\label{fig:chi2cut}}
\end{figure}

A visual investigation of the pulses revealed two different classes of
signal: one that contained one peak, which the functional form
reproduced well; and one that contained multiple peaks, which the
functional form did not reproduce well. The poorly fitted pulses were
clearly due to noise. Cutting on
$\log(\delta^{2})<-4$~(\parfig~\ref{fig:chi2cut}), defined by
\begin{displaymath}
  \delta^2 = h^{-2} \sum_i [V_{i}-f(t_{i})]^{2},\nonumber
\end{displaymath}
rejects such noise, where $V_i$ and $t_i$ are the voltage and time of
the $i$th voltage sample, $f(t_{i})$ is the fitted voltage at time
$t_i$, and $h$ is the maximum voltage of the event.

\subsection{Detector Calibration \label{sec:detcal}}

\begin{figure}[t]
  \centering
  \includegraphics[width=\textwidth]{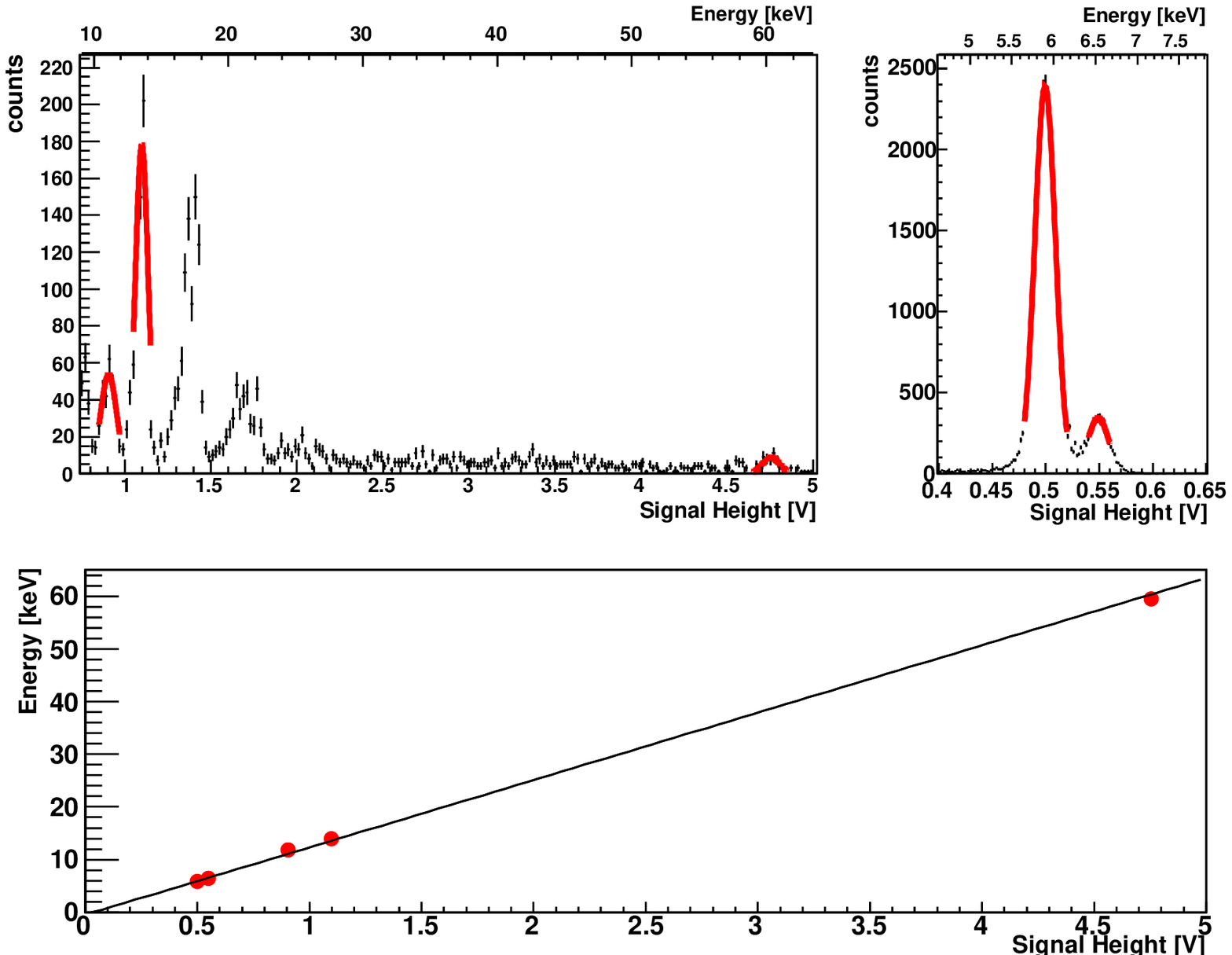}
  \caption{\americium\ (top left) and \iron\ (top right) \xray\
    spectra, and calibration fit (bottom).\label{fig:detcal}}
\end{figure}

To calibrate the detector and readout system, we record the spectra of
the known \xray\ sources \iron\ and \americium. The two sources
provide \xrays\ at many energies between \unit{5}{\keV} and
\unit{60}{\keV}, allowing us to learn the proportionality of the
detectors signal amplitude to~\Tdep~(\parfig~\ref{fig:detcal}) and to
check the linearity of the detector response.

\begin{figure}[t]
  \centering
  \includegraphics[angle=-90,width=.45\columnwidth]{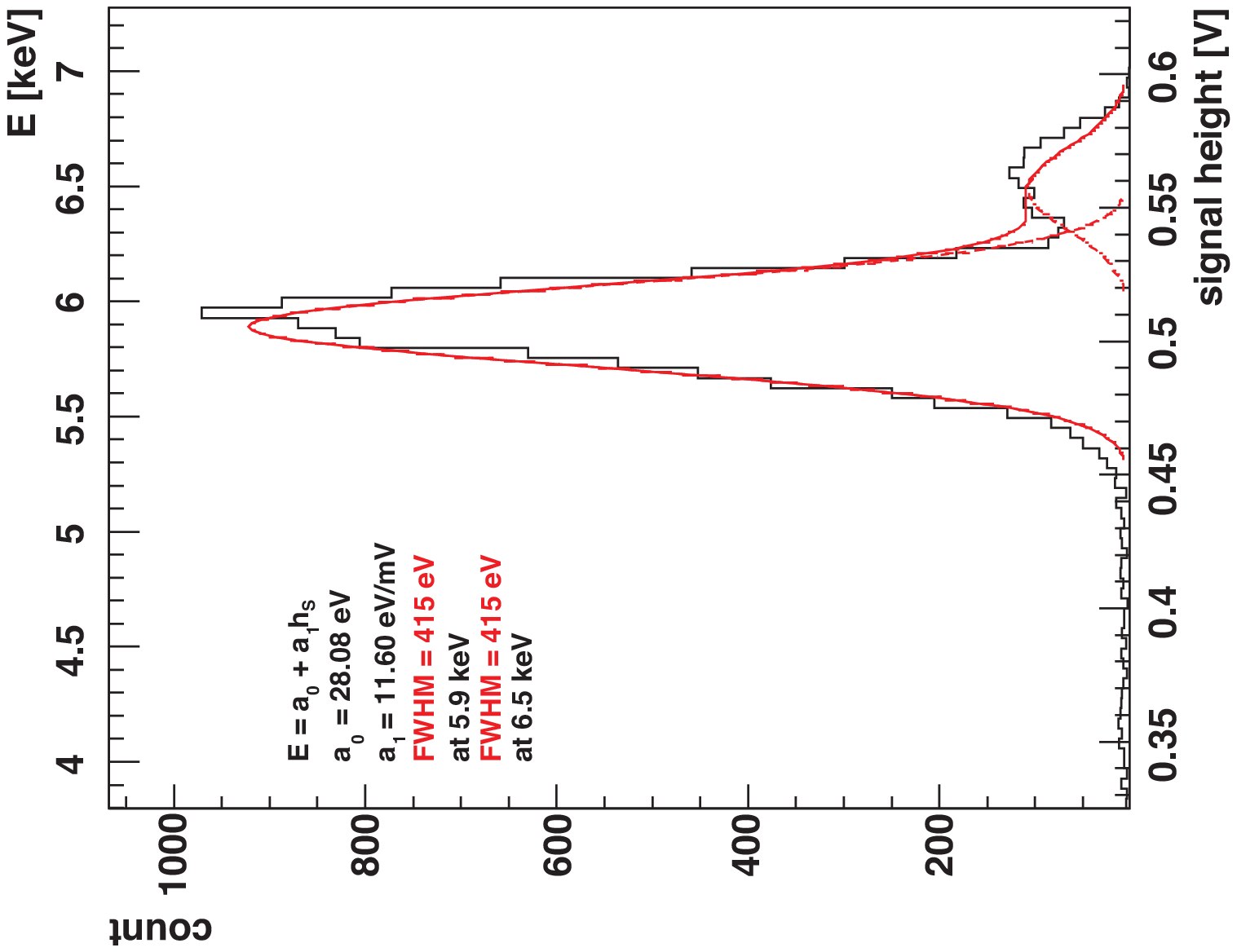}
  \includegraphics[angle=-90,width=.45\columnwidth]{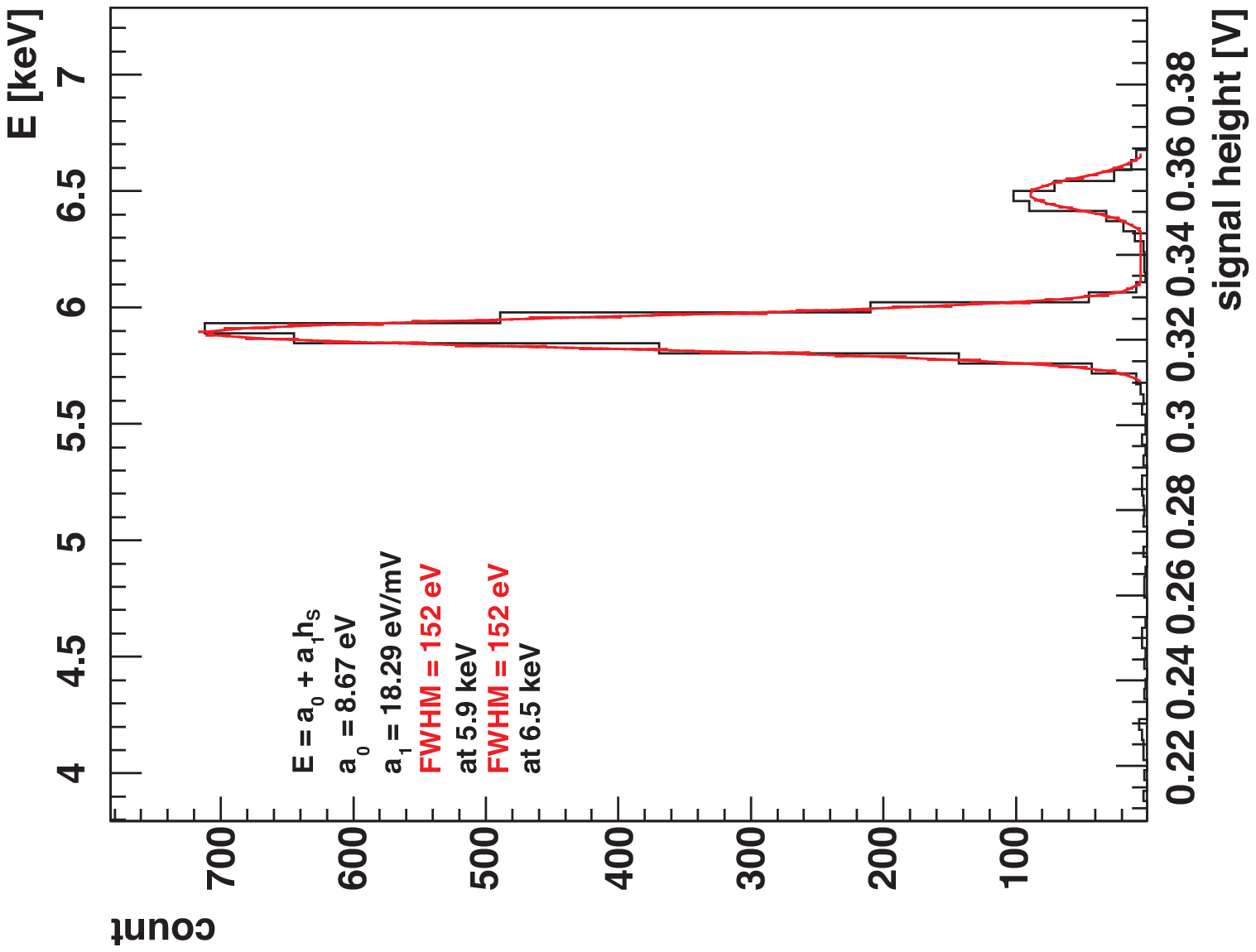}
  \caption{\iron\ spectra taken with the \SDD\ at room temperature
    (left) and cooled to \unit{-20}{\celsius} (right).\label{fig:detres}}
\end{figure}

Since \xrays\ deposit their energy locally and uniformally inside the
silicon, these sources also allow us to measure the energy resolution
of the detector and our read-out system without influences from any
dead layer structure. The resolution, here defined as the \FWHM\ of an
energy peak, is greatly affected by the temperature of the
detector~(\parfig~\ref{fig:detres}).  At room temperature, the
detector has resolutions at the \MnKa\ line of
\iron~(\unit{5.9}{\keV}) greater than~\unit{400}{\eV}. When cooled to
temperatures below~\unit{0}{\celsius}, the detector reaches
resolutions below~\unit{150}{\eV}.

\subsection{Gas Purity \label{sec:gaspurity}}

The frictional cooling effect and the operation of the open \SDD\
require that the moderating gas be uncontaminated.

The detector has a dead layer of \unit{30}{\nano\meter} of aluminum on
top of the silicon that forms its active volume. As well, in the first
approximately \unit{200}{\nano\meter} of the silicon, the detector has
a charge collection efficiency below 100\%. In the energy range of
interest, $T\lessapprox\unit{30}{\keV}$, protons deposit a significant
amount of energy in the aluminum and deposit their remaining energy
largely in the region of reduced collection efficiency.  Therefore,
the \SDD\ measures only 35\% to 65\% of the proton's energy
(\parsec~\ref{sec:protons}).

Impurities in the gas environment can build up on the detector's
surface, which is the coldest surface in the experiment setup,
increasing the amount of material that protons must traverse before
entering the active layers of the detector. Even a thin layer of
built-up gas impurities significantly reduces the amount of energy
measured and, due to straggling in the trajectories of protons through
the dead layers, greatly reduces the detector's energy resolution. The
main sources of impurities are outgassing of molecules from the
plastic pieces in the experiment construction and contamination of the
gas from impurities either in the gas source or entering along the gas
transfer line.  The effects of outgassing can be greatly reduced by
pumping the gas cell down to a low pressure
(\unit{\power{10}{-6}}{\milli\bbar}) for several days. During this
time, the plastic pieces expel foreign molecules, leaving the
environment cleaner for data-taking runs.

The boil-off from cryogenic liquids is used as an ultrapure gas
source. The gas transfer lines are tightly sealed, capable of holding
pressures down to \unit{\power{10}{-9}}{\milli\bbar}, to prevent air
leaking into the gas. As well, the transfer lines are entirely
constructed from stainless steel, so no outgassing can contaminate the
gas on the way from the source to the cell. The construction of an
impurity trap, improvements in the gas line, and ultrapure helium
sources will be implemented for running he full system.

\subsection{Electric Breakdown}

Due to the large electric field strengths it provides, the
accelerating grid must be kept in high vacuum
($P<\unit{\power{10}{-4}}{\milli\bbar}$) in order to prevent breakdown
of the electric field between grid rings. Even at high vacuum,
breakdowns can occur if the lines carrying the high voltage come too
near to lower voltage rings or grounded pieces. Such breakdowns were
observed in early versions of the experiment construction in which the
high voltage was lead to the \HV\ ring by an insulated wire that
passed along the length of he accelerating grid. Electrical discharges
originated at the \HV\ ring and traveled down the surface of the
wire's insulation to the grounded detector flange, when high voltages
as low as \unit{15}{\kV} were applied to the \HV\ ring. To prevent
this from occuring, the construction was altered to bring high
voltages to the \HV\ ring from the opposite side
(\parfig~\ref{fig:fcdphoto})

The gas seal at the source side of the cell must be very tight since
it is at the \HV\ end of the grid. A small leak can stream gas over
the rings creating a path for the breakdown between rings. It can also
provide a path for charge to flow in a breakdown from inside the gas
cell to the \HV\ ring. Both breakdowns mechanisms were observed in an
early version of the experiment construction in which the proton
source was mounted directly through the gas cell wall. This
construction did not create a tight enough gas seal, and breakdowns
were observed at gas pressures above several
\unit{\power{10}{-1}}{\milli\bbar}, when high voltages as low as
\unit{10}{\kV} were applied to the \HV\ ring. To prevent gas leaks
causing such breakdowns, we constructed the source-holder platform
described above.

\subsubsection{Breakdown Inside the Gas Cell}

Breakdown of the electric field also occurs entirely inside the gas
cell. This can be particularly dangerous since a breakdown that
strikes the detector can destroy it or the detector electronics. The
frequency and strength of these breakdowns depends greatly on the
source-holder platform.

We constructed two platforms: one made of \PEEK\ that does not alter
the electric field of the accelerating grid; and one made of stainless
steel that can be connected to the \HV\ ring or left unconnected
(floating), and alters the shape of the electric field at the \HV\
end.

When the metal platform is electrically connected to the \HV\ ring,
such a large current is drawn at any pressure that the high voltage
supply is incapable of providing enough current to hold a \HV\ above
\unit{1}{\kilo\volt}. However, when the platform is left floating,
higher voltages can be reached before a breakdown from the platform to
a point inside the gas cell occurs.

\begin{figure}[t]
  \centering
  \includegraphics[width=\textwidth]{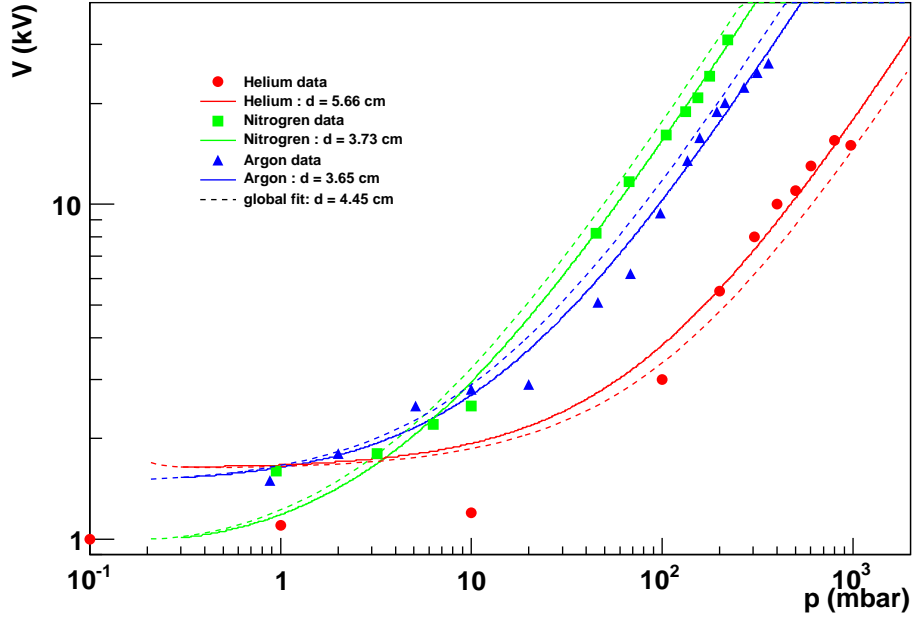}
  \caption{Threshold voltage for breakdown of the electric field
    inside the gas cell for helium, nitrogen, and argon with a
    floating metal source-holder platform.\label{fig:paschen}}
\end{figure}

A photodiode, mounted onto an empty \SDD\ housing that was mounted
inside the gas cell in place of the detector, measured the light from
discharges in the gas allowing for measurement of the frequency of
electric breakdown. We observed that disharges of a harmless size
occured frequently (\unit{\gtrapprox1}{\hertz}), but did not drain
enough current to alter the voltage. However, above a voltage
threshold that depends on the pressure of the gas in the cell, we
observed large discharges occuring with higher frequency. These
discharges prevented the high voltage supply from keeping a constant
voltage. The dependence of the threshold voltage \sub{V}{br} on gas
pressure $p$ (\parfig~\ref{fig:paschen}) matches the predictions from
the Townsend discharge theory \cite{pej2002} augmented by an offset
voltage $V_{0}$,
\begin{equation}
  \sub{V}{br} = \frac{Bpd}{\ln(Apd)-\ln(\ln(1+\gamma^{-1}))} + V_{0},
\end{equation}
where $d$ is the distance over which the discharge takes place, $A$
and $B$ are the Townsend coefficients, and $\gamma$ is the secondary
emission coefficient. Coefficients $A$, $B$, and $\gamma$ are
different for each gas. Fitting the Townsend theory to the
voltage-threshold data with $d$ and $V_0$ as the only free parameters
indicates a discharge distance on the order of \centi\meter, the same
order of size as the gas cell.

%\comment{source vs no source + statement about charge clearing
%  considerations}

Tests conducted with the \PEEK\ platform and americium source indicate
that charge freed from the gas by the americium alphas builds up on
the platform until a breakdown occurs from the platform to the \HV\
ring. These breakdowns create large discharges and are fatal to the
detector and detector electronics. However, these breakdowns do not
occur when the seal between the \PEEK\ platform and the end of the gas
cell is made very gas tight, closing the path for charge to take
during a breakdown. In the gas-tight cell, electric fields of
strengths up to \unit{650}{\kilo\volt\per\meter} have been reached
without breakdown at gas pressures from
\unit{\power{10}{-7}}{\milli\bbar} to \unit{1.25}{\bbar}.

\section{FCD Cell Simulation\label{sec:CoolSim}}

We simulated the frictional cooling process in the \FCD\ cell to
provide expectations for proton energy spectra under different
configurations. For this simulation as well as full frictional cooling
schemes~\cite{greenwald:293,Bao201028}, we developed software based
on Geant4, called CoolSim~\cite{CoolSim}. It implements the low-energy
packages of Geant4 optimized for the tracking of protons and muons
through matter. As well, we have added new processes to the Geant4
framework for the simulation of charge exchange processes at low
energies in gaseous materials \cite{greenwald:293}.

\begin{figure}[t]
  \centering
  \includegraphics[width=\columnwidth]{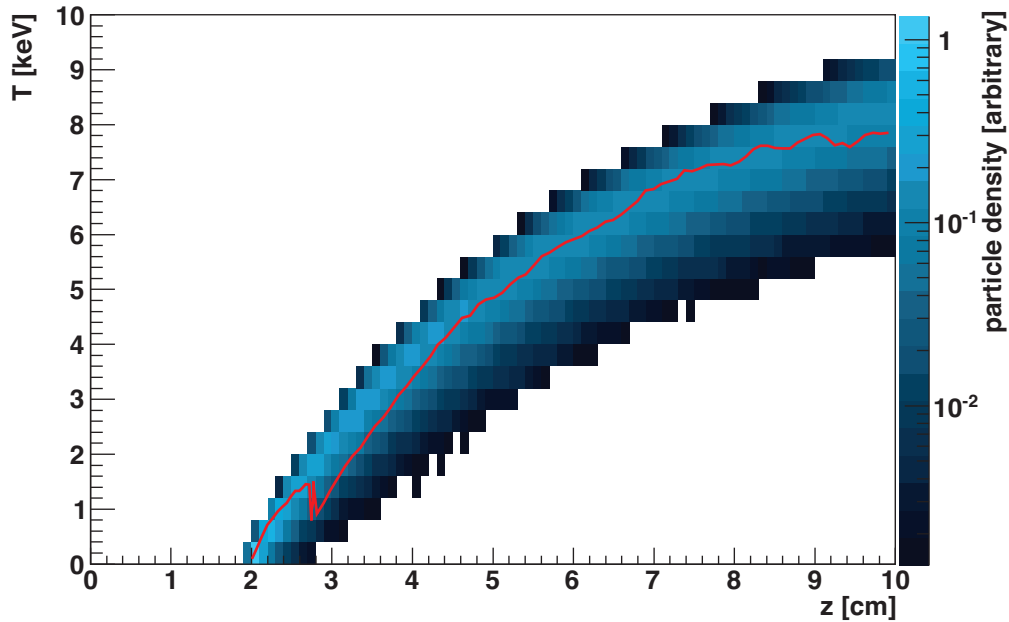}
  \caption{Simulated kinetic energy distributions of protons (shaded)
    and the kinetic energy of a single proton (line) as a function of
    $z$ in the \FCD\ cooling cell filled with helium gas at
    \unit{40}{\milli\bbar} and an electric field strength of
    \unit{0.4}{\MVpm}.\label{fig:evz}}
\end{figure}

The cell simulated was exactly as described in
\textsec~\ref{sec:construction}, with the electric field shape as
calculated in \textsec~\ref{sec:sim:efield}. The protons were
simulated as originating from a point source located on the $z$ axis
at the center of the 5th ring ($z=\unit{20}{\milli\meter}$). In the
following discussion all data are taken from runs in which ten
thousand protons were simulated for each of the combinations of nine
electric field strengths, evenly spaced between \unit{0.1}{\MVpm} and
\unit{0.5}{\MVpm}, and nine helium gas densities, logarithmically
spaced between \unit{1}{\milli\bbar} and \unit{700}{\milli\bbar}. In
each simulation run, protons start at rest and accelerate through the
gas in the positive $z$ direction, approaching the equilibrium energy
(\parfig~\ref{fig:evz}).

\begin{figure}[t]
  \centering
  \includegraphics[width=\columnwidth]{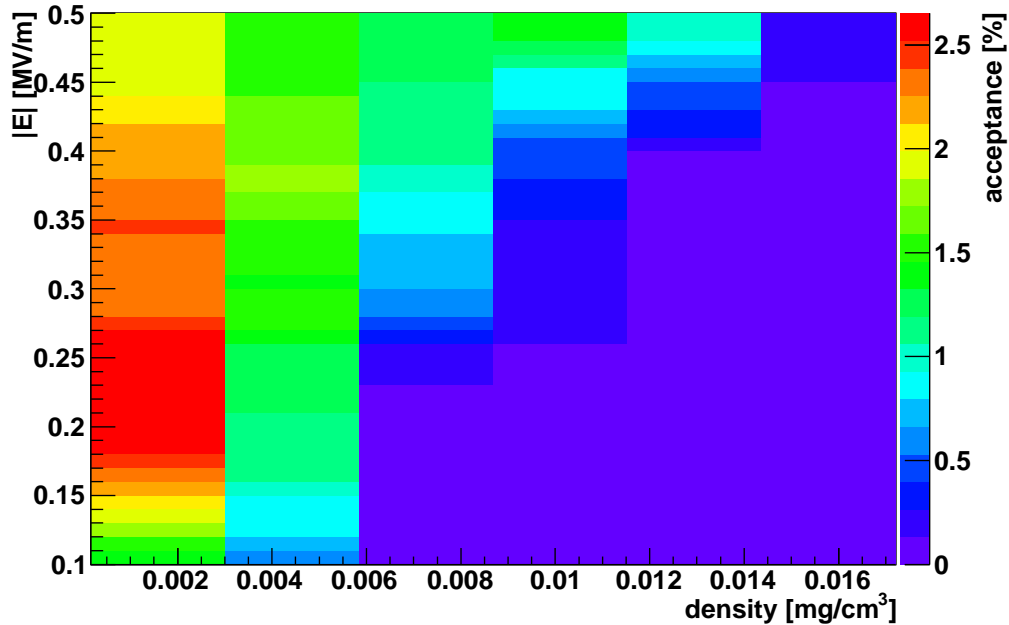}
  \caption{Simulated detector acceptance as a function of electric
    field strength and helium gas density.\label{fig:acceptance}}
\end{figure}

\label{sec:detAcc}
As they accelerate, they interact with the helium gas, scattering away
from the $z$ axis and decreasing acceptance in the
\SDD~(\parfig~\ref{fig:acceptance}), which has a radius of
\unit{1.78}{\milli\meter}. The mean free path for scattering decreases
with increasing helium gas pressure, causing more muons to scatter
away from the \SDD\ at higher pressures. However, the stronger
electric fields refocus some of those scattered protons towards the
$z$ axis. The lowest acceptances are expected for the
high-pressure--weak-field region of the parameter space. The
scattering can be seen in the highlighted trajectory of
figure~\ref{fig:evz}: after scattering into a direction opposed to the
electric field, the proton decelerates, turns around, and then
reaccelerates. This produces the abrupt kinetic energy fluctuation
seen in the figure.

\begin{figure}[t]
  \centering
  \includegraphics[width=\columnwidth]{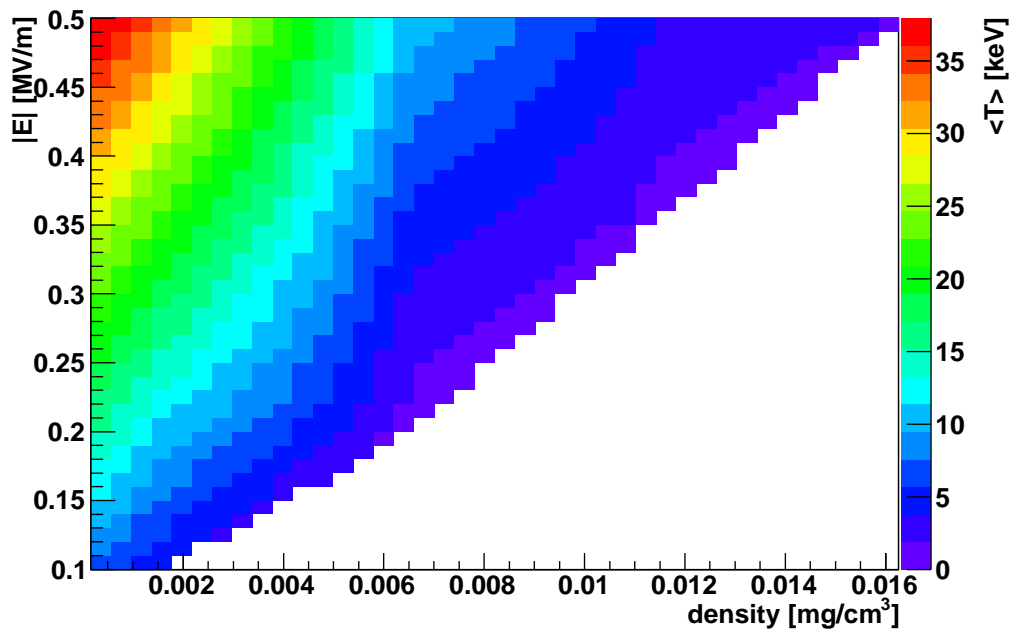}
  \caption{Simulated mean kinetic energy of protons at the \SDD\ as a
    function of electric field strength and helium gas
    density.\label{fig:E_P_T}}
\end{figure}

For each combination of electric field strength and gas pressure the
mean of the kinetic energy distribution at the \SDD\
($z=\unit{10}{\centi\meter}$, $r<=\unit{1.78}{\milli\meter}$) is
calculated (\parfig~\ref{fig:E_P_T}). For a fixed electric field
strength, raising the gas pressure increases the energy loss to the
helium, decreasing the mean energy at the detector. For a fixed gas
pressure, raising the electric field strength increases the
restorative energy gain, increasing the mean energy. Both behaviors
are as expected from \textfig~\ref{fig:dedx}.

\section{Measurements}

Several measurements were made using the experimental setup to
calibrate the detectors~(\parsec~\ref{sec:detcal}) and measure the
effect of their dead layers, as well as to measure the \xray\
background, and verify the production of protons. All of the following
measurements were made with the proton source, with a
$23$-$\micro\meter$-thick Mylar foil, mounted in the accelerating grid
at $z=\unit{20}{\milli\meter}$ and the detector cooled to
approximately \unit{15}{\celsius}. Background and gasless proton
measurements were made without the gas cell in place. The total data
taking rate was approximately \unit{10}{\hertz}. The \xray\ rate was
approximately \unit{1.5}{\hertz}; the proton rate was approximately
\unit{2}{\hertz}; and the remaining rate was due to saturated pulses
from \MeV\ alpha particles.

\subsection{Background}

\begin{figure}[t]
  \centering
  \includegraphics[width=\columnwidth]{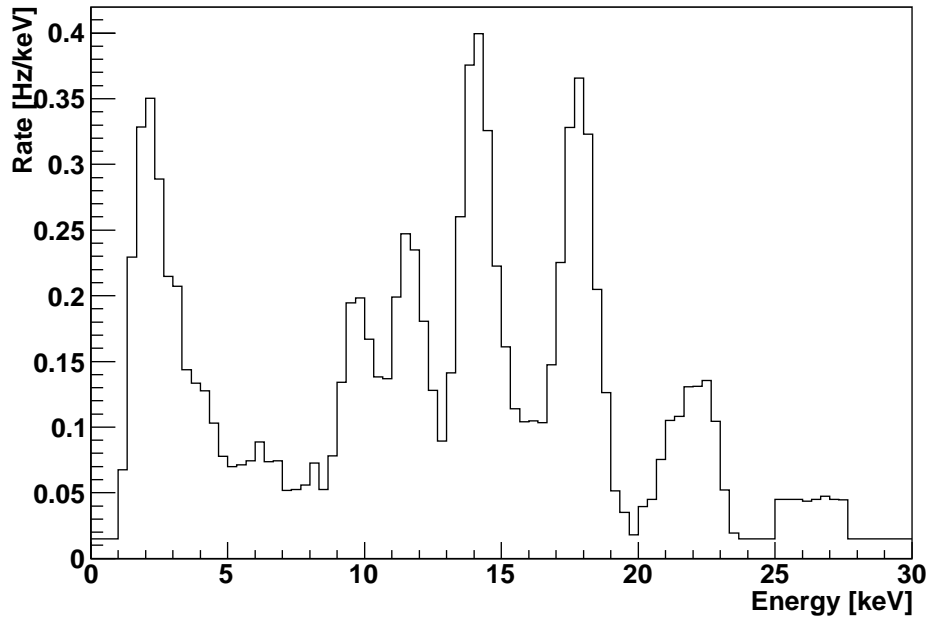}
  \caption{Background spectra with \americium\ source present and the
    gas cell evacuated.\label{fig:background}}
\end{figure}

\begin{table}[t]
  \begin{center}
    \begin{tabular}{r @{.} l r @{.} l r r @{.} l}
      \hline
      \hline

      \multicolumn{2}{l}{\rule{0pt}{1em}Energy} &
      \multicolumn{2}{c}{\ BR} &
      \multicolumn{1}{r@{\hspace{.6em}}}{D} &
      \multicolumn{2}{c}{BR\cdot D} \\

      \multicolumn{2}{l}{(\keV)} &
      \multicolumn{2}{c}{\ (\%)} &
      \multicolumn{1}{r@{\hspace{.25em}}}{\ (\%)} &
      \multicolumn{2}{l}{(\%)} \\

      \hline
      11&87  &  0&66   &  92  &   0&61  \\
      13&76  &  1&07   &  81  &   0&87  \\
      13&95  &  9&6    &  79  &   7&6   \\
      15&86  &  0&15   &  62  &   0&10  \\
      16&11  &  0&18   &  61  &   0&11  \\
      16&82  &  2&5    &  58  &   1&4   \\
      17&06  &  1&5    &  56  &   0&8   \\
      17&50  &  0&65   &  54  &   0&35  \\
      17&99  &  1&37   &  51  &   0&70  \\
      20&78  &  1&39   &  36  &   0&50  \\
      21&10  &  0&65   &  35  &   0&23  \\
      21&34  &  0&59   &  35  &   0&20  \\
      21&49  &  0&29   &  34  &   0&10  \\
      26&34  &  2&40   &  23  &   0&56  \\
      59&54  &  35&90  &  3   &   1&21  \\
      \hline
      \hline
    \end{tabular}
  \end{center}
  \caption{Energies, branching ratios (BR), detectability (D),
    and BR\cdot D for \xrays\ emitted by \americium\ with
    branching ratio greater than 0.2\%\label{tab:AmXrays}}	
\end{table}	

The \americium\ in the proton source emits \xrays\ in the energy range
of interest for the proton measurements ($E\lesssim\unit{30}{\keV}$).
The observed rate of \xrays\ is comparable to that of protons, so the
background energy spectrum must be measured for subtraction from the
proton energy spectra. The background
spectrum~(\parfig~\ref{fig:background}) contains several peaks from
the \americium\ spectrum~(\partab~\ref{tab:AmXrays}) as well as
low-amplitude noise. To reduce the low-amplitude noise rate, a voltage
threshold corresponding to an energy threshold of \unit{1}{\keV} is
used in data recording. The probability that the \xrays\ interact in
the \SDD, which is \unit{450}{\micro\meter} thick, rapidly decreases
with increasing energy in the range of interest.
Table~\ref{tab:AmXrays} lists the detectability (D), defined as the
percentage of \xrays\ interacting in the sensitive volume of the
detector, and detectable branching ratio (D\cdot BR) for the prominent
\xray\ lines.

\subsection{Proton Observations \label{sec:protons}}

\begin{figure}[t]
  \includegraphics[width=\columnwidth]{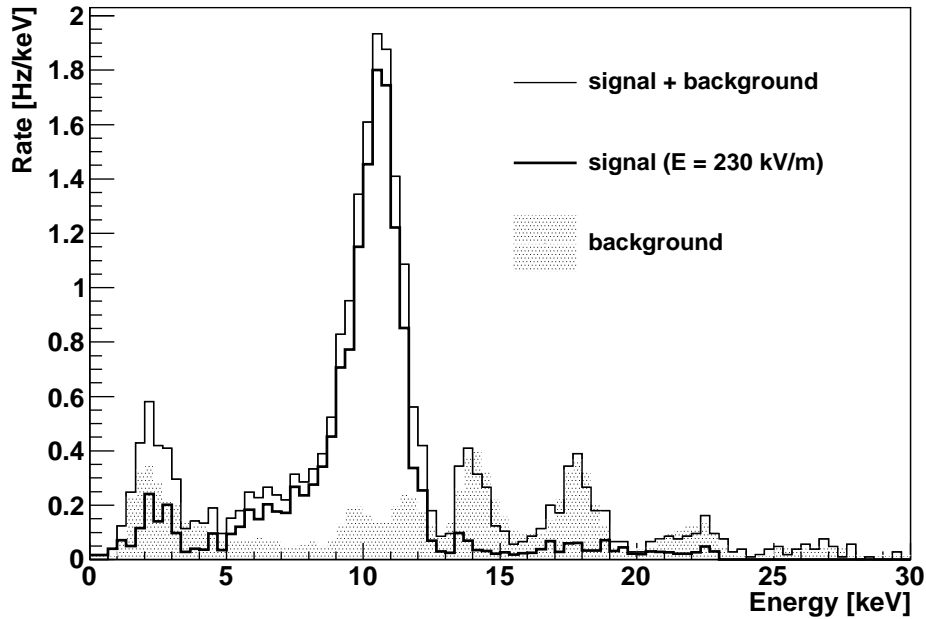}
  \caption{Energy spectrum for an $E=\unit{23}{\kVpm}$ run: full
    spectrum (line), background-removed spectrum (thick line), and
    background (highlighted).\label{fig:bgsubtraction}}
\end{figure}

Energy spectra were measured with electric field strengths evenly
spaced from \unit{70}{\kVpm} to \unit{300}{\kVpm} in \unit{10}{\kVpm}
steps. The background spectrum ($E=\unit{0}{\kVpm}$) and proton
spectra ($E>\unit{0}{\kVpm}$) were analyzed together to discover the
overall background rate and the signal rate above that background for
each spectrum. Figure~\ref{fig:bgsubtraction} shows an example
spectrum, for $E=\unit{230}{\kVpm}$. The background spectrum as
calculated from all the measured spectra is shown for comparison. The
spectrum has a prominent proton peak centered around approximately
\unit{11}{\keV}, which is lower than the \unit{18.4}{\keV} expected
from the calculation of the electric field. The discrepancy is due to
energy deposition in the dead and partially-inactive layers of the
detector. The peak has a \FWHM\ of approximately \unit{2}{\keV}. This
is larger than the \xray\ energy resolution at the same energy, but is
as expected from the distribution of proton energy loss in the dead
layers.

The proton peak also has a tail to lower energies. This is due to
protons striking the outer edges of the detector surface where they
encounter larger dead layers. The increase in the low-energy noise
above the background rate may be due to fluorescence of the silicon
and aluminum of the detector, which produces peaks in this range.

\begin{figure}[t]
  \includegraphics[width=\columnwidth]{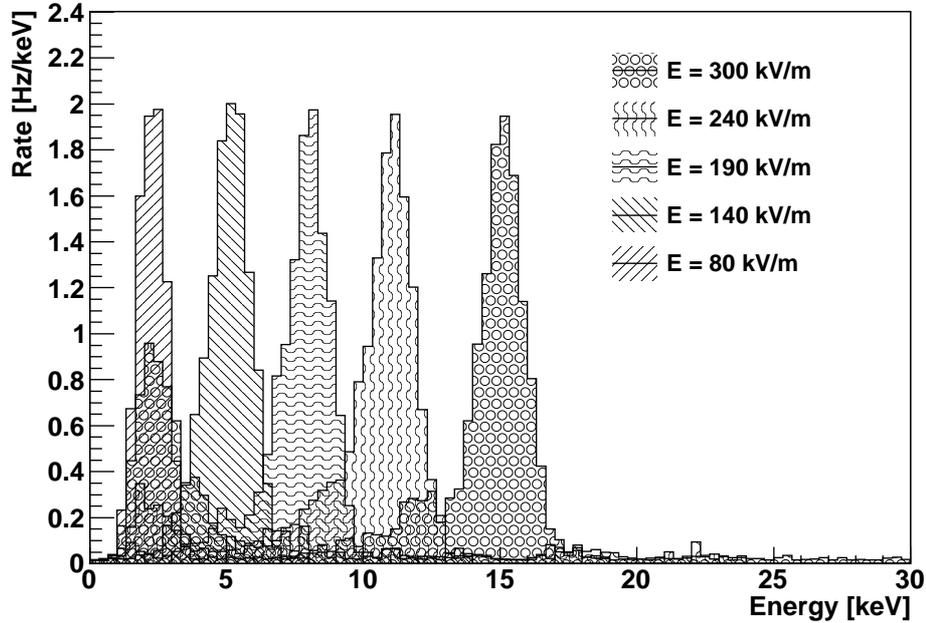}
  \caption{Proton energy spectra, with background removed, for five
    representative strengths of the electric field.\label{fig:protonspec}}
\end{figure}

Figure~\ref{fig:protonspec} shows five of the proton spectra along
with the overall background. The peak centers are evenly spaced in
accordance with the field strengths at which they were measured. As
well, the proton rate remained nearly constant with changing electric
field strength.

\begin{figure}[!t]
  \includegraphics[width=\columnwidth]{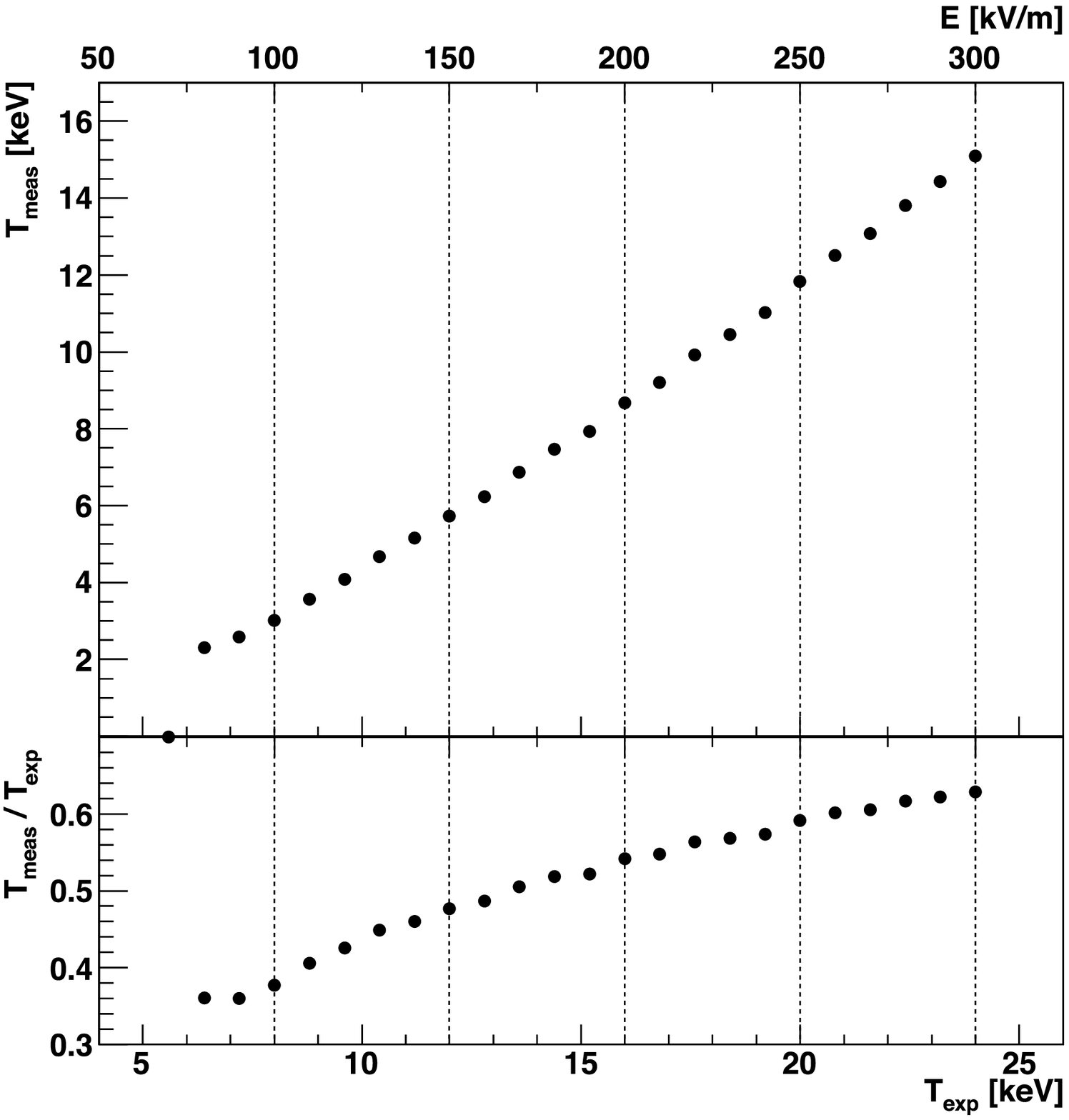}
  \caption{\SDD-measured energy as a function of expected proton
    energy (top) and the ratio of measured to expected energy as a
    function of expected energy (bottom). \comment{combine into one
      plot}\label{fig:detresponse}}
\end{figure}

The upper plot of figure~\ref{fig:detresponse} shows the peak centers
of the spectra (\sub{T}{meas}, obtained by fitting with a gaussian
distribution) as a function of the expected proton energy
(\sub{T}{exp}), which is obtained from from the numerical calculation
of the electric field. The detector's dead layers are of a thickness
on the order of 100s of \nano\meter, which is also the order of size
of the penetration depth of \keV\ protons. The higher-energy protons
travel further into the detector, depositing a larger ratio of their
energy in fully active layers of the detector
(\parfig~\ref{fig:detresponse}, bottom).

The lowest energy run in \textfig~\ref{fig:detresponse}, for which
$E=\unit{70}{\kVpm}$, has $\sub{T}{meas}=\unit{0}{\keV}$ because after
depositing energy in the dead layers, the protons didn't have enough
energy left to be measureable above the low-energy threshold, which
vetoes electronic noise in the detector readout system. Any further
build up of a dead layer increases the minimum energy protons must
have in order to be detectable.

\section{Conclusion}

The \FCD\ experiment at the Max Planck Institute for Physics, Munich,
has been commissioned to study the working principle behind frictional
cooling. The experiment construction is complete and all parts have
been commissioned: The accelerating grid can maintain electric field
strengths without breakdown up to \unit{900}{\kVpm} in an evacuated
gas cell and up to \unit{650}{\kVpm} in a pressurized gas cell with
pressures up to \unit{1.25}{\bbar}. The detector can measure energy
with good resolutions and can be reliably operated in the strong field
of the accelerating grid. The gas system is capable of maintaining a
specified pressure for several hours. Proton spectra have been
measured, demonstrating that the source functions. The next step in
the \FCD\ experiment is the taking of data with the gas cell filled.

\section*{Acknowledgements}
Design and contruction of the experiment apparati was accomplished
with help from Karlheinz Ackermann and G\"unter Winkelmuller. Many of
the data-taking electronics components were designed and built by Si
Tran. Much help in operating the \SDD s and their control electronics
was given by Adrian Niculae and Atakan Simsek from PNSensor.
Commissioning and data taking was accomplished with the help of
Christian Blume, Raphael Galea, Andrada Ianus, Brodie Mackenzie, Alois
Kabelschacht, and Franz Stelzer.

%\bibliographystyle{model1-num-names}
%\bibliography{fcd.bib}

\end{document}